\documentclass[11pt]{article}

\newsavebox{\foobox}
\newcommand{\slantbox}[2][0]{\mbox{%
        \sbox{\foobox}{#2}%
        \hskip\wd\foobox
        \pdfsave
        \pdfsetmatrix{1 0 #1 1}%
        \llap{\usebox{\foobox}}%
        \pdfrestore
}}
\newcommand\unslant[2][-.25]{\slantbox[#1]{$#2$}}

\newcommand{\mpi}{\text{\unslant[-.18]\pi}}
\newcommand{\mdelta}{\text{\unslant[-.18]\delta}}

\usepackage[left=2cm, right=2cm, top=2.5cm, bottom=2.5cm]{geometry}
\usepackage[x11names]{xcolor}
\usepackage{fancyhdr, amssymb, cancel, amsmath, graphicx, pgfplots, tikz}
\usepackage{isomath}

\usetikzlibrary{shadows}

\newcommand{\stylecolor}{IndianRed3}

\usepackage[labelfont={bf,sf, color=\stylecolor}, margin={1.5cm,0cm}]{caption}

\usepackage[colorlinks=true, urlcolor=\stylecolor, linkcolor=\stylecolor, citecolor=\stylecolor, hyperindex=true, linktocpage=true]{hyperref}

\usepackage{amsthm}
\newtheoremstyle{theor}{10pt}{10pt}{}{16pt}{\sffamily \bfseries \color{green!50!black}}{:}{.5em}{}
\theoremstyle{theor}

\usepackage[explicit]{titlesec}

\newcommand*\sectionlabel{}
\titleformat{\section}
  {\gdef\sectionlabel{}
   \Large\bfseries\scshape}
  {\gdef\sectionlabel{\thesection }}{0pt}
  {\begin{tikzpicture}[remember picture,overlay]
       \end{tikzpicture}
  }
\titlespacing*{\section}{0pt}{0pt}{0pt}

\newcommand*\subsectionlabel{}
\titleformat{\subsection}
  {\gdef\subsectionlabel{}
   \large\bfseries\scshape}
  {\gdef\subsectionlabel{\thesubsection  }}{0pt}
  {\begin{tikzpicture}[remember picture]
    	\draw (-0.15, 0) node[left] {\color{\stylecolor} \textsf{\subsectionlabel}};
	\draw (0.15, 0) node[right] {\color{\stylecolor} \textsf{#1}};
	\fill[color=\stylecolor] (-0.05, -0.23) rectangle (0.05, 0.23);
       \end{tikzpicture}
  }
\titlespacing*{\subsection}{-4pt}{10pt}{0pt}

\newcommand*\subsubsectionlabel{}
\titleformat{\subsubsection}
  {\gdef\subsubsectionlabel{}
   \bfseries\scshape}
  {\gdef\subsubsectionlabel{\thesubsubsection.\ \  }}{0pt}
  {\begin{tikzpicture}[remember picture]
    	\draw (0, 0) node[left] {\color{\stylecolor} \textsf{\subsubsectionlabel}};
	\draw (0, 0) node[right] {\color{\stylecolor} \textsf{#1}};
       \end{tikzpicture}
  }
\titlespacing*{\subsubsection}{-4pt}{7pt}{0pt}

\pgfplotsset{every axis legend/.append style={at={(1.02,1)},anchor=north west}}

\pgfplotsset{every axis legend/.append style={at={(1.02,1)},anchor=north west}}

\begin{document}

%\allowdisplaybreaks

\pagestyle{fancy}
\renewcommand{\headrulewidth}{0pt}
\fancyhead{}

\fancyfoot{}
\fancyfoot[C] {\textsf{\textbf{\thepage}}}

\begin{equation*}
\begin{tikzpicture}
\draw (\textwidth, 0) node[text width = \textwidth, right] {\color{white} easter egg};
\end{tikzpicture}
\end{equation*}

\begin{equation*}
\begin{tikzpicture}
\draw (0.5\textwidth, -3) node[text width = \textwidth] {\huge  \textsf{\textbf{Constraints on hydrodynamics from many-body \\ \vspace{0.07in} quantum chaos}} };
\end{tikzpicture}
\end{equation*}
\begin{equation*}
\begin{tikzpicture}
\draw (0.5\textwidth, 0.1) node[text width=\textwidth] {\large \color{black} \textsf{Andrew Lucas}};
\draw (0.5\textwidth, -0.5) node[text width=\textwidth] {\small \textsf{Department of Physics, Stanford University, Stanford, CA 94305, USA}};
\end{tikzpicture}
\end{equation*}
\begin{equation*}
\begin{tikzpicture}
\draw (0, -13.15) node[right, text width=0.5\paperwidth] { \texttt{ajlucas@stanford.edu}};
\draw (\textwidth, -13.1) node[left] {\textsf{\today}};
\end{tikzpicture}
\end{equation*}
\begin{equation*}
\begin{tikzpicture}
\draw[very thick, color=\stylecolor] (0.0\textwidth, -5.75) -- (0.99\textwidth, -5.75);
\draw (0.12\textwidth, -6.25) node[left] {\color{\stylecolor}  \textsf{\textbf{Abstract:}}};
\draw (0.53\textwidth, -6) node[below, text width=0.8\textwidth, text justified] {\small  
Is the hydrodynamics of an interacting many-body system fundamentally limited by basic principles of quantum mechanics?  Starting with the conjecture that viscosity is at least as large as entropy density (as measured in fundamental units), there has been a long search for a precise answer to this question.    In this work, we identify a simple relationship between hydrodynamics and many-body quantum chaos in a broad class of experimentally realizable systems.  Consistency with the quantum butterfly effect leads to upper bounds on the speed of sound and diffusion constants of hydrodynamics.    These bounds link two very different theories of quantum many-body dynamics, clarify the relationship between classical hydrodynamics and quantum information loss,  and provide a simple way to constrain theories of thermalization and quantum chaos in experiments.
 };
\end{tikzpicture}
\end{equation*}

\tableofcontents

\begin{equation*}
\begin{tikzpicture}
\draw[very thick, color=\stylecolor] (0.0\textwidth, -5.75) -- (0.99\textwidth, -5.75);
\end{tikzpicture}
\end{equation*}

\titleformat{\section}
  {\gdef\sectionlabel{}
   \Large\bfseries\scshape}
  {\gdef\sectionlabel{\thesection }}{0pt}
  {\begin{tikzpicture}[remember picture]
	\draw (0.2, 0) node[right] {\color{\stylecolor} \textsf{#1}};
	\fill[color=\stylecolor]  (0,0.37) rectangle (-0.7, -0.37);
	\draw (0.0, 0) node[left, fill=\stylecolor] {\color{white} \textsf{\sectionlabel}};
       \end{tikzpicture}
  }
\titlespacing*{\section}{0pt}{20pt}{5pt}

\section{Introduction}
The time evolution of a many-body quantum system is unitary, and thus reversible.  Yet the classical world we experience is dissipative and irreversible.  A simple way to understand this irreversibility is as follows.  If there are $\mathcal{K}$ interacting $N$-state systems, there are $N^{\mathcal{K}}$ possible states that the many-body wave function could be in.   No realistic observer or computer can hope to keep track of so much information in the thermodynamic limit $\mathcal{K} \rightarrow \infty$.   Hence, a typical observer, with access to a finite number $\mathcal{M} \ll \mathcal{K}$ of degrees of freedom (DOF), will always lose track of any perturbations to a typical excited state.   The spreading of quantum information from $\mathcal{M}$ degrees of freedom to the remaining $\mathcal{K}-\mathcal{M}$ degrees of freedom is what ultimately allows the observer to perceive the state as thermal.  It is an acknowledgement that there is too much quantum information to keep track of \cite{sekino}.

Many-body quantum chaos \cite{bhbutterfly} is the theory of this scrambling of quantum information.  An important prediction of this theory  is that in a spatially extended system, quantum information does not spread instantaneously \cite{localized, lrbutterfly, nahum}.   An observer with access to all degrees of freedom in a region of length $L$ will only lose track of quantum information after a time $t\sim L/v_{\mathrm{B}}$,  where $v_{\mathrm{B}}$ is the butterfly velocity.    $v_{\mathrm{B}}$ often depends on the state of the quantum system, but not  on $L$.   

After quantum information has been sufficiently lost, an observer may see hydrodynamic evolution of the system.  Hydrodynamics is the effective  theory describing the relaxation of an interacting many-body system to global thermal equilibrium \cite{kadanoff}.   If $\rho$ is a conserved density, its expectation value $\langle \rho(x,t)\rangle$ becomes well approximated by the solution to a classical differential equation with suitable initial conditions.  Typically, the density propagates ballistically in a sound wave:
 \begin{equation}
\partial_t^2 \langle \rho \rangle \approx v_{\mathrm{s}}^2 \partial_x^2 \langle \rho \rangle + \Gamma_{\mathrm{s}} \partial_t\partial_x^2 \langle \rho \rangle,   \label{eq:sound}
\end{equation} or spreads diffusively: \begin{equation}
\partial_t \langle \rho \rangle = D \partial_x^2 \langle \rho\rangle,  \label{eq:fick}
\end{equation}
or a combination of both.   The coefficients $D$ and $v_{\mathrm{s}}$ have been measured in a diverse family of experimental systems, including cold atomic gases \cite{cao}, quark-gluon plasma \cite{shuryak}, and strange metals \cite{ramshaw}.   Hydrodynamics gives a simple, tractable limit in which the response of strongly coupled theories is strongly constrained.   Understanding the hydrodynamic limit well is necessary  to understand the more general problem of quantum many-body dynamics.

Unfortunately, diffusion constants are notoriously hard to compute from first principles in strongly interacting many-body systems.  Based on the notion that the ``interaction time" cannot be faster than $\hbar/k_{\mathrm{B}}T$ \cite{sachdev, zaanen}, and  conjectures that conductivities and viscosities are fundamentally bounded from below \cite{kss, ritz},  \cite{hartnoll2014} proposed that Planckian scattering  bounds diffusion:  \begin{equation}
D \gtrsim v^2 \frac{\hbar}{k_{\mathrm{B}}T}.  \label{eq:Dinit}
\end{equation}
A remarkable proposal \cite{blakeB1, blakeB2} was that the velocity scale in this conjectured bound should be  $v_{\mathrm{B}}$, and much evidence for such a relation has been subsequently put forth \cite{aleiner, gu, patel, erez, blakeB3, gu2, mendl, ehud, patel2, kkim1, matteo, blakeB4, kkim2}.   While (\ref{eq:Dinit}) is known to have counterexamples \cite{julia, gouteraux, gu2017chain}, one could ask whether chaos does bound diffusion in some other way.

Because hydrodynamics is a dissipative theory describing the evolution of local observables, and chaos describes the process by which quantum systems appear thermal to local observers, hydrodynamics should be constrained by the spread of many-body quantum chaos.  For example, a dissipative sound wave cannot propagate faster than the speed of decoherence and scrambling;  hence we conjecture that \begin{equation}
v_{\mathrm{s}} \le v_{\mathrm{B}}.  \label{eq:mainv}
\end{equation}
Similarly,  dissipative diffusive spreading cannot surpass the light cone where quantum information is scrambled.   Because diffusion occurs whenever $x^2 \lesssim Dt$, if hydrodynamics is valid only for times $t\ge \tau$, we conclude that $(v_{\mathrm{B}}\tau)^2 \gtrsim D\tau$ in order for quantum information to be scrambled before diffusion takes place.  Thus, we conjecture that \begin{equation}
D\le v_{\mathrm{B}}^2 \tau.  \label{eq:mainD}
\end{equation}
The logic above has recently been used to bound $D \le v_{\mathrm{LR}}^2 \tau$ \cite{hartnoll1706}, where  $v_{\mathrm{LR}}$ is the Lieb-Robinson velocity: a state-independent velocity describing the growth of initially local operators  \cite{liebrobinson}.    In many systems, $v_{\mathrm{B}} \ll v_{\mathrm{LR}}$ \cite{lrbutterfly}.   The bound (\ref{eq:mainD}) can be parametrically stronger than the bound proposed in \cite{hartnoll1706}.

We will make the conjectures (\ref{eq:mainv}) and (\ref{eq:mainD}) precise in theories with $N\sim 1$ DOF per  lattice site, starting from very simple assumptions.    When $N\gg 1$, there are multiple velocities associated with quantum information scrambling.   We will describe how (\ref{eq:mainv}) and (\ref{eq:mainD}) generalize to such theories, and  resolve multiple paradoxes in large $N$ theories where hydrodynamics appears faster than scrambling.

\section{Small $N$ Models}
Let us first focus on a many-body quantum system with $N\sim1$ DOF per lattice site.   We define the butterfly velocity as the smallest velocity for which the inequality \begin{align}
\left| \left\langle  A(x,t) [B(x,t),C(0)]D(0)  \right\rangle_\beta    \right|, \left| \left\langle [C(0),  A(x,t)] B(x,t)D(0)  \right\rangle_\beta \right|  \le  \frac{1}{N} \mathcal{G}\left(|x|-v t\right).
\label{eq:vB}
\end{align} 
holds, with $v=v_{\mathrm{B}}$.   Here, $A$, $B$, $C$ and $D$ are local, non-Hermitian operators, $\langle \cdots \rangle_\beta = Z(\beta)^{-1} \mathrm{tr}[\rho_\beta \cdots ]$ is the thermal expectation value at temperature $T=1/\beta k_{\mathrm{B}}$, $\mathcal{G}(0)$ is an O(1) number and $\mathcal{G}(x)$ decays faster than $|x|^{-2-d}$ as $x\rightarrow +\infty$, with $\mathcal{G}(\infty)=0$.    $v_{\mathrm{B}}$ is the speed with which local operators at spacetime $(0,0)$ no longer commute with operators at $(v_{\mathrm{B}}t,t)$, in thermal expectation values.   If the commutator (\ref{eq:vB}) approximately vanishes, then in typical homogeneous quantum systems one expects \cite{bhbutterfly, localized} \begin{equation}
\langle A(x,t)B(x,t)C(0)D(0)\rangle_\beta \approx \langle A(x,t)C(0)B(x,t)D(0)\rangle_\beta \approx \langle A(0)B(0)\rangle_\beta\langle C(0)D(0)\rangle_\beta,  \label{eq:2com}
\end{equation}
and observers at $(x,t)$ are unaffected by perturbations at $(0,0)$.  However, when $|x|\ll v_{\mathrm{B}}t$, the two correlation functions in (\ref{eq:2com}) are not the same:  in many instances the out-of-time-ordered correlator will vanish \cite{bhbutterfly, localized}.   The deviation of these two correlation functions tells us that quantum information has scrambled.     The postulate of many-body chaos is that in typical ergodic systems, a velocity $v$ exists for which the bound (\ref{eq:vB}) is reasonably tight (at least for large enough $x$).  At short times, the chaotic lightcone may not be sharp \cite{nahum}, and one can generalize (\ref{eq:vB}) by replaceing $\mathcal{G}(|x|-vt)$ with $\mathcal{G}(|x|-\int \mathrm{d}t v(t))$, where $v(t)$ is the velocity with which information propagates at time $t$.

In many lattice models, the conserved density $\rho$ may be written as a composite operator
\begin{equation}
\rho(x,t) = \sum_I \Phi^\dagger_I(x,t) \Phi_I(x,t)   \label{eq:density}
\end{equation}
where $\Phi^\dagger_I$/$\Phi_I$  are creation/annihilation operators.\footnote{The index $I$ can be a sum over vector indices, or a matrix trace, depending on the nature of the degrees of freedom, but we define $N$ such that $I=1,\ldots,N$.}     Hydrodynamic operators often schematically take the form of (\ref{eq:density}), though they may involve more operators than just $\Phi_I/\Phi_I^\dagger$.  
For simplicity, we assume $\rho$ is defined in (\ref{eq:density}) for now, leaving the generic case to Appendix \ref{app:multi}.   Our essential arguments can be understood from this simple example.

The retarded Green's function \begin{equation}
G^{\mathrm{R}}_{\rho\rho}(x,t) = \mathrm{i}\mathrm{\Theta}(t)\langle [\rho(x,t),\rho(0,0)]\rangle_\beta  \label{eq:GR}
\end{equation}
takes a specific form in the hydrodynamic regime.  Using (\ref{eq:density}), we can relate this two-point function to four-point functions of $\Phi^\dagger_I/\Phi_I$:\begin{align}
G^{\mathrm{R}}_{\rho\rho}(x,t) &= \mathrm{i\Theta}(t)\sum_{IJ}\langle \Phi^\dagger_I(x,t)[\Phi_I(x,t),\Phi^\dagger_J(0,0)]\Phi_J(0,0)+ [\Phi^\dagger_I(x,t), \Phi^\dagger_J(0,0)] \Phi_I(0,0) \Phi_J(x,t)+  \notag \\
&+ \Phi^\dagger_J(0,0)\Phi^\dagger_I(x,t)[\Phi_I(x,t),\Phi_J(0,0)] + \Phi^\dagger_J(0,0)[\Phi^\dagger_I(x,t),\Phi_J(0,0)] \Phi_I(x,t)  \rangle_\beta .  \label{eq:vBineq}
\end{align}
Thus, \begin{align}
\left|G^{\mathrm{R}}_{\rho\rho}(x,t)\right| \le \sum_{IJ} \left|\langle \Phi^\dagger_I(x,t)[\Phi_I(x,t),\Phi^\dagger_J(0,0)]\Phi_J(0,0) \rangle_\beta \right| + \text{3 similar terms}.  \label{eq:vBineq2}
\end{align}
From (\ref{eq:vB}) we obtain  \begin{equation}
\left|G^{\mathrm{R}}_{\rho\rho}(x,t)\right| \le 4N \mathcal{G}(|x|-vt). \label{eq:GRvB}
\end{equation}
with $v=v_{\mathrm{B}}$ when $N\sim 1$.  $G^{\mathrm{R}}_{\rho\rho}$ is suppressed whenever $|x|\ge v_{\mathrm{B}} t$.   

If $\rho$ has overlap with only a diffusive hydrodynamic mode at late times, in $d$ spatial dimensions, \cite{kadanoff} \begin{equation}
G^{\mathrm{R}}_{\rho\rho} = -D\chi \nabla^2 \frac{\mathrm{e}^{-x^2/4Dt}}{(4\mpi Dt)^{d/2}},  \label{eq:diffineq}
\end{equation}
whenever $t\ge \tau$;  $\chi$ is a thermodynamic prefactor.  Up to the overall $\nabla^2$, (\ref{eq:diffineq}) is just the kernel of the classical differential equation (\ref{eq:fick}).  $G^{\mathrm{R}}_{\rho\rho}$ is not small when $x^2 \lesssim Dt$, so consistency with (\ref{eq:GRvB}) requires $4\mathcal{G}(\sqrt{Dt} - v_{\mathrm{B}}t) \gtrsim D\chi (Dt)^{-1-d/2}$.   Since $\mathcal{G}$ rapidly vanishes at large $x$, (\ref{eq:diffineq}) is only valid for time $t>\tau$, where $\sqrt{D\tau} \lesssim v_{\mathrm{B}}\tau$.   Giving a precise definition for $\tau$ and $\mathcal{G}$, one can prove an exact bound (\ref{eq:mainD}): see Appendix \ref{app:formal}.  In what follows, we will only compute $\tau$ up to an O(1) constant, as in \cite{hartnoll1706}:  this is sufficient  to understand a set of qualitative tensions between diffusion and chaos that follow.

If $\rho$ has overlap with a ballistic hydrodynamic mode, then at late times the Green's function $G^{\mathrm{R}}_{\rho\rho}$ is given (up to total $x$-derivatives) by a $d$-dependent function which is not exponentially suppressed at the sound front $|x| = v_{\mathrm{s}}t$.   Since $\mathcal{G}$ is rapidly suppressed whenever $|x|\ge v_{\mathrm{B}}t$,  we derive (\ref{eq:mainv}).   The argument of the previous paragraph  also implies $\Gamma_{\mathrm{s}} \le v_{\mathrm{B}}^2 \tau$, where  $\Gamma_{\mathrm{s}}$ controls the diffusive spreading of sound waves in (\ref{eq:sound}).

Recent numerics show that $v_{\mathrm{B}}$ can be mildly sensitive to whether or not we take $\mathcal{G}(0)$ to be parametrically small, when measured in microscopic units \cite{mendl}.  However, in a typical small $N$ theory, the hydrodynamic susceptibilities such as $\chi$ are not parametrically small.  Therefore, a significant fraction of scrambling must take place, and the commutator in (\ref{eq:vB}) must be reasonably large, before hydrodynamics is consistent with (\ref{eq:GRvB}).   The butterfly velocity bounds hydrodynamics.  

 In  theories of interest including superfluids \cite{putterman}, quantum-fluctuating superconductors \cite{hartnollSC} or Fermi liquids with exotic band structure \cite{hartnoll1704, lucas1706}, there may be more conservation laws than energy, momentum and charge.  In this case, the Green's function of the diffusive conserved quantities $G^{\mathrm{R}}_{\rho_a \rho_b}$ obeys a matrix-valued generalization of the Fourier transform of (\ref{eq:diffineq}), with indices running over $ab$;  similar comments hold when there are multiple sound modes.   The derivations above hold for every diffusion constant, and every sound mode.

If $v(t)$ in (\ref{eq:vB}) is taken to be time-dependent, then (\ref{eq:mainv}) generalizes to $v_{\mathrm{s}} \le v(\infty)$ while (\ref{eq:mainD}) generalizes to $D \le \tau^{-1}[\int_0^\tau \mathrm{d}t v(t)]^2$.   While many simple thermalizing systems have constant $v(t)$, in many-body localized states  $v(t) \lesssim 1/t$ \cite{xie2017, fan2017out, chenlogarithm, debanjan2}.  Applying (\ref{eq:mainD}) at late times one finds $D=v_{\mathrm{s}}=0$ \cite{hartnoll1706}.

  \section{Large $N$ Models}
  We now turn to theories with $N\rightarrow \infty $ local DOF.    At  strong coupling, many such theories are observed to have OTOCs of the form \cite{bhbutterfly, localized, aleiner, gu, patel,   erez, stanfordbound, douglasweak, debanjan} \begin{equation}
  \left| \left\langle  A(x,t) [B(x,t),C(0)]D(0)  \right\rangle_\beta    \right| \le  \frac{\mathrm{e}^{(t-|x|/v_{\mathrm{B}})/\tau_{\mathrm{L}}}}{N} +  \frac{a(x,t)}{N} + \mathrm{O}\left(\frac{1}{N^2}\right)  \label{eq:vBreal}
  \end{equation}
  with $a(x,t)$ a bounded function obeying $a(\infty,t)=0$.
  Following the convention in the literature \cite{mezei}, we will define the butterfly velocity $v_{\mathrm{B}}$, at large $N$, to be  the velocity associated with the exponentially growing light cone in (\ref{eq:vBreal}).   Simply put, $v_{\mathrm{B}}$ is the speed with which all $N$ DOF can begin to scramble.  
  
\begin{figure}
\centering
\includegraphics{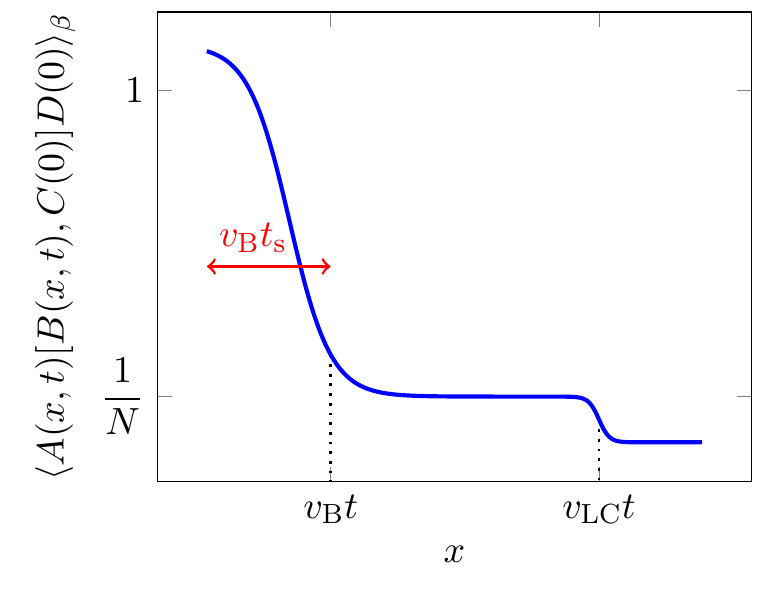}
\caption{A sketch of the possible behavior of the OTOC in a large $N$ system.   The OTOC saturates to $\mathrm{O}(1/N)$ at the light cone $x=v_{\mathrm{LC}}t$ -- this is sufficient scrambling for hydrodynamic behavior.  However, most information may only scramble at the butterfly lightcone $x=v_{\mathrm{B}}t$.  Once the butterfly lightcone has passed, it takes a scrambling time $t_{\mathrm{s}}\sim \tau_{\mathrm{L}} \log N$ for the OTOC to saturate, assuming (\ref{eq:vBreal}). }
\label{fig:main}
\end{figure}

In a typical large $N$ theory,  thermodynamic data such as $\chi \propto N$, while diffusion constants and sound speeds $\propto N^0$.    From the derivation of (\ref{eq:GRvB}) from (\ref{eq:vB}), it is clear that the OTOCs of $\Phi^\dagger_I/\Phi_I$ need only be $\sim 1/N$ in order for $G^{\mathrm{R}}_{\rho\rho}\propto N$ to be possible.    Therefore, we define from (\ref{eq:vB}) a light cone velocity, $v=v_{\mathrm{LC}}$, where $\mathcal{G}(0)$ is O(1) (and not $\mathrm{O}(N)$) for typical operators: see Figure \ref{fig:main}.  $v_{\mathrm{LC}}$ sets the speed with which a few DOF have scrambled.  The bounds (\ref{eq:mainv}) and (\ref{eq:mainD}) therefore become \begin{subequations}\label{eq:either}\begin{align}
v_{\mathrm{s}} &\le v_{\mathrm{LC}}, \\
D &\le v_{\mathrm{LC}}^2 \tau.
\end{align}\end{subequations}    

When $t\gg |x|/v_{\mathrm{B}}$, and exponentially fast scrambling is occuring at a fixed $x$,  the triangle inequality employed in (\ref{eq:GRvB}) becomes very weak:  the right hand side $\propto N^2$, even as $G^{\mathrm{R}}_{\rho\rho} \propto N$.   Parametric cancellations between the OTOCs in (\ref{eq:vBineq}) must occur at large $N$ for $t>|x|/v_{\mathrm{B}}$.   The growth rate of chaos, the Lyapunov time $\tau_{\mathrm{L}}$, plays no role in (\ref{eq:mainD}).    The scrambling of a few DOF is sufficient to constrain hydrodynamics.  This is the key difference between finite $N$ and large $N$ theories.
The link between chaos and hydrodynamics is clearly strongest when $N$ is small, and $v_{\mathrm{B}}\approx v_{\mathrm{LC}}$ cannot parametrically be distinguished.

Nevertheless, much of the evidence for a connection between the butterfly velocity and diffusion comes from large $N$ theories.   In what follows, we will briefly discuss and resolve some existing tensions between hydrodynamics and quantum chaos in strongly coupled large $N$ theories, leaving technical calculations to the appendices.  For simplicity, we set $\hbar=k_{\mathrm{B}}=1$ henceforth.

The charged Sachdev-Ye-Kitaev (SYK) chain  \cite{gu2} is a solvable chaotic lattice model of $N\gg 1$ fermions coupled together via random quartic couplings.  Charge and energy are locally  conserved, the charge density operator is given by (\ref{eq:density}), and one can explicitly compute the functional form of the OTOCs (\ref{eq:vB}):  see  Appendix \ref{app:syk}.  One finds that \cite{gu, gu2} hydrodynamics is valid for times $t \gg 1/T$, and that the heat diffusion constant $D_{\mathrm{h}} = v_{\mathrm{B}}^2/2\mpi T$ ``saturates" (\ref{eq:mainD}).   The charge diffusion constant $D_{\mathrm{c}}$ is not universally linked to $D_{\mathrm{h}}$.  If $D_{\mathrm{c}}\gg D_{\mathrm{h}}$ were possible,  we observe directly from OTOCs that $v_{\mathrm{LC}}$ beomes time-dependent while $v_{\mathrm{B}}$ is time-independent: $v_{\mathrm{LC}}>v_{\mathrm{B}}$ at early times, and $D_{\mathrm{c}} \sim v_{\mathrm{LC}}(\beta)^2\beta$.    However, in analytically tractable limits we always find $D_{\mathrm{c}}<D_{\mathrm{h}}$.   This suggests that charge diffusion does not generically precede the butterfly  light cone: $v_{\mathrm{LC}}=v_{\mathrm{B}}$, and (\ref{eq:mainD}) is obeyed.

Another class of QFTs where we may reliably compute $D$ and $v_{\mathrm{s}}$, together with $v_{\mathrm{B}}$ and $v_{\mathrm{LC}}$, are large $N$ theories holographically dual to gravity in one higher dimension \cite{lucasrmp, koenraadbook}.   A direct comparison of our formalism to holographic models is not easy: we do not know how to directly write hydrodynamic operators $\rho$ in the form of (\ref{eq:density}).   We postulate that (\ref{eq:either})  continues to hold in these models.
A puzzling feature of many holographic models is that $v_{\mathrm{s}}$ often appears to be much larger than $v_{\mathrm{B}}$.     Yet sound waves can only propagate after (some) information has scrambled.  
One might guess that \begin{equation}
v_{\mathrm{LC}} = \max(v_{\mathrm{s}},v_{\mathrm{B}}). \label{eq:VLC}
\end{equation} 
To confirm this, we must compute OTOCs holographically by studying the gravitational response of a two-sided black hole to a small amount of spatially localized infalling matter \cite{localized}:  see Appendix \ref{app:holovB}.   Infalling matter becomes exponentially blueshifted as it gets close to the black hole horizon, and as a consequence a  spatially localized gravitational shockwave forms.   This gravitational shockwave is responsible for the first  term of (\ref{eq:vBreal}), which (at $N=\infty$) grows unboundedly in time.  However, the infalling matter will also generically excite fluctuations of the geometry which are dual to sound waves.   The resulting metric fluctuations are not blueshifted at late times, and are responsible for $a(x,t)$ in (\ref{eq:vBreal}).  We find that $a(x,t)$ does not vanish when $|x|=v_{\mathrm{s}}t$.   We conclude that $v_{\mathrm{B}}$ continues to be set by the gravitational shockwave,  while $v_{\mathrm{LC}}$ may be fixed by sound.
There is no contradiction with (\ref{eq:mainv}).

We now study holographic charge-neutral plasmas.   The charge density $\rho$ decouples from energy-momentum dynamics and diffuses at late times.   
In some models $D\sim v_{\mathrm{B}}^2 /T$ \cite{blakeB1}, which is suggestive of (\ref{eq:Dinit}) with Planckian scattering $\tau \sim 1/T$.   However, other models have $DT/v_{\mathrm{B}}^2$ unboundedly large \cite{blakeB1}.   We conjecture that (\ref{eq:fick}) breaks down at a time $\tau \gtrsim D/v_{\mathrm{B}}^2$, even though  $\tau$ is very large compared to the Planckian scattering time $1/T$.  
 On first glance, this is surprising.  Holographic models are strongly interacting and one typically expects to see hydrodynamic behavior on time scales $t\gg 1/T$ \cite{lucasrmp}.
To test this conjecture, we must determine how (\ref{eq:fick}) can break down.  Either a non-hydrodynamic excitation has a lifetime $\tau$ \cite{hartnoll1706}, and/or higher derivative corrections to (\ref{eq:fick}) become important at time $t=\tau$.   We first estimate the lifetime of the non-hydrodynamic modes, which holographically are dual to quasinormal modes of a black hole.   We show in Appendix \ref{app:holoC} that in general, the lifetime of quasinormal modes can be very short compared to $D/v_{\mathrm{B}}^2$.    Next, we compute higher derivative corrections to (\ref{eq:fick}):  see Appendix \ref{app:holoC}.  Such corrections appear to become non-negligible on length scales $\ell \gtrsim D/v_{\mathrm{B}}$, or on time scales $\tau \gtrsim D/v_{\mathrm{B}}^2$.  This is precisely the time scale where the diffusion of charge is contained within the butterfly lightcone.

In theories with a conserved momentum and charge/energy,  the speed of sound is fixed by thermodynamics.  The DOF responsible for sound must scramble at this speed, while the rest scramble far slower.   The SYK chain and holographic charge-neutral plasma can both be studied without momentum conservation.    There is no obvious symmetry or conservation law allowing any DOF to scramble faster than any other.   Perhaps this is why we find  evidence for the stronger bound (\ref{eq:mainD}) in both of these large $N$ models.

\section{Outlook}
The quantitative link between quantum many-body chaos and hydrodynamics was first noticed in \cite{blakeB1, blakeB2}.  In this letter, we have presented a simple physical picture linking these two very different theories of many-body dynamics.     
Our constraints on hydrodynamics lead to upper bounds, in contrast to the conjecture (\ref{eq:Dinit}).  Neither (\ref{eq:mainD}) nor previous lower bounds on transport \cite{lucas, grozdanov, grozdanov2} formally invoke a bound on $\tau_{\mathrm{L}}$ \cite{stanfordbound}, or the Heisenberg uncertainty principle, which provides an informal bound $\tau \gtrsim \hbar/k_{\mathrm{B}}T$.   Whether the Heisenberg uncertainty principle leads to any formal bound on transport coefficients in a many-body quantum system remains an important open problem. 

We have observed that sufficient scrambling of quantum information, \emph{but not} the spreading of quantum entanglement, appear prerequisite to the onset of hydrodynamics.  In large $N$ models, entanglement and chaos can spread  with parametrically different velocities \cite{mezei2}.  It would be interesting to find a lattice model with $N\sim 1$, together with hydrodynamics outside of an ``entanglement lightcone".

Direct measures of OTOCs are challenging \cite{monika, grover, yao}, though it is possible in small systems \cite{garttner, du, cappellaro}.  But in experimentally realized systems with $N\sim 1$, the bounds (\ref{eq:mainv}) and (\ref{eq:mainD}) directly relate the quantum butterfly effect to sound and diffusion, which can be measured directly in many-body systems.  Non-trivial insight and constraints on the  spread of thermalization and chaos may  be possible with simple hydrodynamic measurements.

\addcontentsline{toc}{section}{Acknowledgments}
\section*{Acknowledgments}

I thank Richard Davison, Raghu Mahajan, Adam Nahum, Sid Parameswaran, Xiao-Liang Qi, Subir Sachdev, Douglas Stanford and especially Mike Blake, Yingfei Gu and Sean Hartnoll, for very useful feedback, and the Aspen Center for Physics, which is supported by Grant PHY-1607611 from the National Science Foundation, for hospitality as this work was begun.    I am supported by the Gordon and Betty Moore Foundation's EPiQS Initiative through Grant GBMF4302.

\begin{appendix}

\section{Generic Hydrodynamic Operators}\label{app:multi}

In this appendix we discuss how the discussions of the main text generalize to more complicated hydrodynamic operators than (\ref{eq:density}).  The most general form for a hydrodynamic singlet operator such as the energy density $\epsilon$ is\footnote{We will use $\epsilon$ in this Appendix, as opposed to $\rho$ in the main text, to minimize confusion.  But we note that in general, the operator $\rho$ will not be as simple as (\ref{eq:density}), but can be expressed analogously to (\ref{eq:densityA}).} \begin{equation}
\epsilon = N \sum_R \frac{1}{|R|}  \sum_{I=1}^{|R|} M^{\alpha\beta}_R \mathcal{O}_{R,I}^\alpha \mathcal{O}^\beta_{R,I}  \label{eq:densityA}  
\end{equation}
where $R$ denotes representations (not necessarily irreducible) of the symmetry group $G$,  $\mathcal{O}^\alpha_{R,I}$ denotes operator $\alpha$ in representation $R$ carrying ``flavor" index $I$,  and $M^{\alpha\beta}_R$ is a matrix dependent on $R$ but not $I$.   It is useful to perform a singular value decomposition on $M_R$: \begin{equation}
M^{\alpha\beta}_R \equiv  \sum_i  U^{\alpha i}_R f_R^i V^{i\beta}_R
\end{equation}
where $i$ sums over any non-zero singular values.  Defining $\mathcal{O}^\alpha_{R,I} U^{\alpha i}_R \equiv \mathcal{O}^U_{R,I}$, and similarly defining $\mathcal{O}^V_{R,I}$, we obtain \begin{equation}
\epsilon =  N \sum_R \frac{1}{|R|} \sum_{I\in |R|} f^i_R  \mathcal{O}^{U_i}_{R,I} \mathcal{O}^{V_i}_{R,I} .  \label{eq:SVD}
\end{equation} 

We define a light cone velocity for the OTOC of four operators, following (\ref{eq:vB}):\begin{equation}
\left|\left\langle \mathcal{O}_{R,I}^{U}(x,t) [\mathcal{O}_{R,I}^{V}(x,t), \mathcal{O}_{R^\prime,I^\prime}^{U^\prime}(0,0)]\mathcal{O}_{R^\prime,I^\prime}^{V^\prime}(0,0)\right\rangle_\beta\right|  \le \frac{1}{N}\mathcal{G}\left[|x|-v_{\mathrm{LC}}(U,V,U^\prime,V^\prime)t\right].  \label{eq:vBA1}
\end{equation}
%If the large $N$ theory is a gauge theory, then we may re-define the butterfly velocity in a gauge invariant (albeit slightly weaker) way: \begin{align}
%\sum_{I,I^\prime} \frac{1}{|R||R^\prime|} &\left|\left\langle \mathcal{O}_{R,I}^{U}(x,t) [\mathcal{O}_{R,I}^{V}(x,t), \mathcal{O}_{R^\prime,I^\prime}^{U^\prime}(0,0)]\mathcal{O}_{R^\prime,I^\prime}^{V^\prime}(0,0)\right\rangle_\beta\right|  \notag \\
%&\le \frac{\mathcal{P}(x,t)}{N}\exp\left[\frac{1}{\tau_{\mathrm{L}}(U,V,U^\prime,V^\prime)} \left(t-\frac{|x|}{v_{\mathrm{LC}}(U,V,U^\prime,V^\prime)}\right)\right].  \label{eq:vBA2}
%\end{align}
Using (\ref{eq:vBA1}) we may now straightforwardly generalize the derivation of (\ref{eq:GRvB}) to obtain (\ref{eq:either}), with the caveat that \begin{equation}
v_{\mathrm{LC}} = \max_{UV,U^\prime V^\prime \in \epsilon} (v_{\mathrm{LC}}(U^i_R,V^i_R,U_{R^\prime}^j, V^j_{R^\prime})).
\end{equation}
We emphasize that the tighest bounds (\ref{eq:either}) are only obtained upon using the smallest possible light cone velocity: in particular, one should compute the maximal value of $v_{\mathrm{LC}}$ only for the operators $U,V,U^\prime,V^\prime$ which can be found in (\ref{eq:densityA}).    When $N \sim 1$, we are free to call the light cone velocity the butterfly velocity, as in the main text.

\section{A Formal Definition of $\tau$} \label{app:formal}
In this appendix, we suggest a formal definition for $\tau$, such that (\ref{eq:mainD}) holds.  The logic is identical to \cite{hartnoll1706}, but we keep explicit track of O(1) prefactors.   We stress that this appendix is included to make our work precise, and to make the derivation of (\ref{eq:mainD}) completely rigorous.   We expect that a heuristic definition of $\tau$, as in \cite{hartnoll1706}, suffices for most practical purposes.   In what follows, we take $v=v_{\mathrm{LC}}$ in (\ref{eq:vB}) and assume that \begin{equation}
\mathcal{G}(x) =\frac{d}{2(2\mpi)^{d/2} \ell^{d+2}\kappa_0} \mathrm{e}^{-x/v_{\mathrm{LC}}\tau_{\mathrm{LC}}}.\label{eq:appBP}
\end{equation}
$\kappa_0$ and $\ell$ are constants;  we will define $\kappa_0$ in (\ref{eq:formalfick}).

Consider the function \begin{equation}
\mathcal{K}(x,t; D) \equiv \left(\frac{d}{2Dt} - \frac{x^2}{(2Dt)^2}\right) \frac{\mathrm{e}^{-x^2/4Dt}}{(4\mpi Dt)^{d/2}},
\end{equation}
which is the retarded Green's function (up to a thermodynamic constant prefactor) of the diffusion equation (\ref{eq:fick}) in $d$ spatial dimensions, computed using the prescription of \cite{kadanoff}.
 We then define $\tau_0$ as the minimal time such that \begin{equation}
\left| \kappa_0 G^{\mathrm{R}}_{\rho\rho}(x,t) - \mathcal{K}(x,t;D)\right| \le \frac{2\mpi d}{(4\mpi Dt)^{1+d/2}}\delta, \;\;\; \text{ for all }   t\ge \tau_0.     \label{eq:formalfick}
\end{equation}
where $\delta \ll 1$ is a user-specified constant, which we will constrain below, and $\kappa_0$ is a dimensional prefactor chosen so that in the late time limit when $G^{\mathrm{R}}_{\rho\rho}$ is diffusive, this inequality is obeyed.   This is the sense in which we mean, in the main text, that Fick's law of diffusion is well obeyed.    It is important that the right hand side of (\ref{eq:formalfick}) decays with the correct power of time $t$;  otherwise such an inequality may become trivial or impossible to satisfy near $x=0$.

We may now prove our bound on $D$.  Let us look for the line $x(t)$ where $\kappa_0 G^{\mathrm{R}}_{\rho\rho}$ is so small, that even if it saturates (\ref{eq:GRvB}) it is just large enough to equal the right hand side of (\ref{eq:formalfick}):
\begin{equation}
\kappa_0 G^{\mathrm{R}}_{\rho\rho} \le \frac{d}{2(2\mpi)^{d/2} \ell^{d+2}} \exp\left[\frac{v_{\mathrm{LC}}t-|x(t)|}{v_{\mathrm{LC}}\tau_{\mathrm{LC}}}\right] = \frac{2\mpi d}{(4\mpi Dt)^{1+d/2}}\delta  \label{eq:xoftineq}
\end{equation}
In this equation, we have taken the liberty of defining a constant $\ell$ such that the first inequality of this equation is equivalent to (\ref{eq:GRvB}).  (\ref{eq:xoftineq}) is obeyed for \begin{equation}
|x| \ge x(t) = v_{\mathrm{LC}}t + v_{\mathrm{LC}}\tau_{\mathrm{LC}} \log \left(\frac{1}{\delta}\left(\frac{\sqrt{Dt}}{\ell}\right)^{d+2} \right).  \label{eq:absx}
\end{equation}
Next, if $x_*$ obeys the inequality \begin{equation}
x_* \ge a_d\sqrt{D\tau_0},  \label{eq:ad}
\end{equation}
with $a_d \ge \sqrt{d/2}$ a constant that depends only on $d$, then we may write \begin{equation}
|\mathcal{K}(x_*,\tau_0;D)| \ge \frac{2\mpi d}{(4\mpi D\tau_0)^{1+d/2}} \exp\left[-\frac{x_*^2}{2D\tau_0}\right].  \label{eq:calGbound}
\end{equation}
From (\ref{eq:formalfick}) and the triangle inequality, we conclude that for all $|x| \ge x(t)$ and $t\ge \tau_0$, we must find \begin{equation}
2\delta \ge \exp\left[-\frac{x^2}{2Dt}\right].  \label{eq:deltabound0}
\end{equation} 
Demanding consistency of (\ref{eq:ad}) leads to
\begin{equation}
\delta \le \frac{1}{2}\mathrm{e}^{-a_d^2/2}.  \label{eq:deltabound}
\end{equation}
Plugging in $x(t)$ into (\ref{eq:deltabound0}), we obtain \begin{equation}
D \le v_{\mathrm{LC}}^2 \tau_0 \dfrac{\displaystyle \left(1+\frac{\tau_{\mathrm{LC}}}{\tau_0} \log\left(\frac{1}{\delta}\left(\frac{\sqrt{D\tau_0}}{\ell}\right)^{d+2}\right) \right)^2}{\displaystyle 2\log\frac{1}{2\delta}}.  \label{eq:Drig}
\end{equation}

It is instructive to think about the limit of $\delta \rightarrow 0$.  In this case, $\tau_0 \rightarrow \infty$.  To see why, consider a generic fourth order correction to the diffusion equation, as given in the dispersion relation (\ref{eq:omegataylor}).  We find that 
\begin{equation}
\kappa_0G^{\mathrm{R}}_{\rho\rho}(0,t) = \mathcal{G}(0,t) \left[ 1 + \frac{(d+2)(d+4)}{4} \frac{r_4}{Dt}  + \cdots \right]
\end{equation}
Hence, we conclude from (\ref{eq:formalfick}) that $\tau_0 \sim \delta^{-1}$ as $\delta \rightarrow 0$.   An optimal bound will generally occur at finite $\delta$.    Picking the larger of the two coefficients in the numerator (\ref{eq:Drig}), we may rigorously write \begin{equation}
D \le  \dfrac{2v_{\mathrm{LC}}^2 \tau_0}{\displaystyle \log\frac{1}{2\delta}} \max \left(1, \left(\frac{\tau_{\mathrm{LC}}}{\tau_0} \log\left(\frac{1}{\delta}\left(\frac{\sqrt{D\tau_0}}{\ell}\right)^{d+2}\right)\right)^2\right).  \label{eq:Dsplit}
\end{equation}
If the first term is larger, then we have a definition of $\tau$;  if the second is larger we need to do a bit more work.  We may re-write the second of the two bounds as \begin{equation}
\frac{\sqrt{D\tau_0}}{\ell} \le (d+2) \sqrt{\dfrac{2}{\displaystyle \log\frac{1}{2\delta}}} \frac{v_{\mathrm{LC}}\tau_0}{\ell} \log \left(\frac{1}{\delta^{1/(d+2)}} \frac{\sqrt{D\tau_0}}{\ell}\right).  \label{eq:Dugly}
\end{equation}

We now must solve an algebra problem.  If \begin{equation}
ax \le \log (bx),  \label{eq:axlogbx}
\end{equation}
then can we find a simple bound for $x>0$ directly?   Rewriting (\ref{eq:axlogbx}) as \begin{equation}
\mathrm{e}^{ax} \le bx \le \frac{2b}{a}\left(\mathrm{e}^{ax/2}-1\right),  \label{eq:eax}
\end{equation}
we conclude that $x \le x_0$ satisfies (\ref{eq:axlogbx}), where $x_0$ is the larger of the two solutions to (\ref{eq:eax}): \begin{equation}
\mathrm{e}^{ax_0/2} = \frac{b}{a} + \sqrt{\frac{b^2}{a^2}-\frac{2b}{a}} < \frac{2b}{a}.  \label{eq:eax0}
\end{equation}

Combining (\ref{eq:Dsplit}) and (\ref{eq:Dugly}) with (\ref{eq:eax0}) we obtain (\ref{eq:mainD}) with  \begin{equation}
\tau \equiv \min_{\delta\ge \exp[-a_d^2/2]/2} \left\lbrace \max\left( \dfrac{2\tau_0(\delta)}{\displaystyle \log\frac{1}{2\delta}}, \frac{\tau_{\mathrm{LC}}^2}{\tau_0(\delta)} \left[ \dfrac{4(d+2)}{\displaystyle \log\frac{1}{2\delta}} \log \left(\frac{(d+2)v_{\mathrm{LC}}\tau_0(\delta)}{\delta^{1/(d+2)}\ell} \sqrt{\dfrac{8}{\displaystyle \log\frac{1}{2\delta}}}\right)  \right]^2   \right) \right\rbrace .
\end{equation}
This is our formal definition of $\tau$ -- for emphasis we have explicitly noted that $\tau_0$ depends on $\delta$.  Clearly such an expression is not particularly elegant.   The key points are as follows:  (\emph{i}) if $\tau_0 \gg \tau_{\mathrm{LC}}$, then one should minimize $\delta$, and then $\tau = 4a_d^{-2} \tau_0  \sim \tau_0$;   (\emph{ii}) if $\tau_{\mathrm{LC}} \sim \tau_0$,\footnote{We do not expect $\tau_{\mathrm{LC}} \gg \tau_0$ to be possible on physical grounds.} then $\tau \sim \tau_{\mathrm{LC}}$ with a complicated constant prefactor.  However, because the constants are inside of a logarithm, such constants are unlikely to be of importance in most theories.   

\section{SYK Chains}\label{app:syk}
Let us provide a few more details of the calculation of (\ref{eq:vB}) in the SYK chain models.  For simplicity, we only discuss the charged SYK chain model explicitly, which has Hamiltonian \cite{gu2}
 \begin{equation}
H = \sum_x  \sum_{I,J,K,L=1}^N  \left[ J_{IJKL,x} \Phi^\dagger_{I,x}\Phi^\dagger_{J,x} \Phi_{K,x} \Phi_{L,x} + J^\prime_{IJKL,x} \Phi^\dagger_{I,x+1}\Phi^\dagger_{J,x+1} \Phi_{K,x} \Phi_{L,x}\right] + \mathrm{H.c.},  \label{eq:sykH}
\end{equation}
where $\lbrace \Phi^\dagger_{I,x}, \Phi_{J,y}\rbrace = \mdelta_{IJ}\mdelta_{xy}$ are conventional fermionic creation/annihilation operators, and $J_{IJKL, x}$ and $J^\prime_{IJKL,x}$ Gaussian random variables of zero-mean and variances \begin{equation}
\overline{J^2_{IJKL,x}} = \frac{4J_0^2}{N^3}, \;\;\;\; \overline{J^{\prime2}_{IJKL,x}} = \frac{4J_1^2}{N^3}.
\end{equation}
See \cite{sachdevye, sachdev98, georges, kitaev, stanford1604} for earlier work on single-site versions of (\ref{eq:sykH}).   In the limit $1\ll \beta J \ll N$, where $J = \sqrt{J_0^2+J_1^2}$, this model becomes strongly coupled and maximally chaotic.  It is this regime which we focus on below.   Following \cite{gu2} we define the charge density 
\begin{equation}
\mathcal{Q} = \frac{1}{N}\sum_{I=1}^N \langle \Phi^\dagger_I \Phi_I\rangle_\beta - \frac{1}{2}.
\end{equation}
There is a generalization of (\ref{eq:sykH}) to couplings between $q$ fermions ($q$ must be  even), and all models for $q\ge 4$ have the same low energy effective field theory.

One can show that the dominant connected contribution to four point correlation functions of the fermions $\Phi$ is $\mathrm{O}(1/N)$.  In Euclidean time $\tau$, and on long length scales compared to the lattice spacing $|x-y|\gg 1$, one finds \cite{gu, gu2} 
\begin{align}
 \frac{1}{N^2}\sum_{I,J=1}^N \langle \Phi_I^\dagger(x,\tau_1)&\Phi_I(x,\tau_2) \Phi^\dagger_J(y,\tau_3)\Phi_J(y,\tau_4)\rangle_\beta \notag \\
&\approx \int \frac{\mathrm{d}p}{2\mpi} C_{\mathrm{h}} \sum_{|n|\ge 2} \frac{\mathrm{e}^{-2\mpi \mathrm{i}n \tau/\beta}}{|n|(n^2-1)(2\mpi T|n| + D_{\mathrm{h}}p^2)} f_{\mathrm{h},n}\left(\frac{\tau_1-\tau_2}{\beta}\right)f_{\mathrm{h},n}\left(\frac{\tau_3-\tau_4}{\beta}\right) \notag \\
&+ C_{\mathrm{c}} \sum_{|n|\ge 1} \frac{\mathrm{e}^{-2\mpi \mathrm{i} n\tau/\beta}}{|n|(2\mpi T|n| + D_{\mathrm{c}}p^2)} f_{\mathrm{c},n}\left(\frac{\tau_1-\tau_2}{\beta}\right)f_{\mathrm{c},n}\left(\frac{\tau_3-\tau_4}{\beta}\right) \label{eq:longsyk}
\end{align}
where $\tau = \frac{1}{4}(\tau_1+\tau_2+\tau_3+\tau_4)$,  $C_{\mathrm{c,h}}$ are dimensionful constant prefactors,  $D_{\mathrm{c,h}}$ are the diffusion constants: 
\begin{subequations}\begin{align}
D_{\mathrm{h}} &=  \frac{v_{\mathrm{B}}^2}{2\mpi T}, \\
D_{\mathrm{c}} &=  \frac{v_{\mathrm{B}}^2}{2\mpi T} \tilde{\gamma},
\end{align}\end{subequations}
where the constant $\tilde\gamma$ admits a large $q$ expansion \begin{equation}
\tilde{\gamma} = \frac{24(1-4\mathcal{Q}^2)}{q^2}  + \mathrm{O}\left(\frac{1}{q^3}\right) \label{eq:tildegamma}
\end{equation}
and $f_{\mathrm{c,h}}$ are functions whose explicit forms, given in \cite{gu, gu2}, are not important for us.     We have neglected lattice effects, which is reasonable in the low temperature limit.    The $C_{\mathrm{h}}$ term of (\ref{eq:longsyk}) is responsible for the exponentially rapid growth of  chaos \cite{gu}.   This is why, as noted in the main text, the heat diffusion constant and butterfly velocity are related.   

Explicit evaluation of OTOCs can be carried out by a careful analytic continuation of this Euclidean time formula to Lorentzian time,  using suitable imaginary regulators to enforce the out-of-time-ordering \cite{roberts}.   Directly carrying out this procedure on the 8 four-point functions shown in (\ref{eq:vBineq}) is tedious;  however, it is also unnecessary for our purposes.   The key observation about (\ref{eq:longsyk}) is that it is, up to overall rescalings of $C_{\mathrm{c}}$ and $C_{\mathrm{h}}$, invariant under the following two rescalings:   \begin{subequations}\begin{align}
t &\rightarrow \lambda t, \;\;\; \beta \rightarrow \lambda \beta, \;\;\; D \rightarrow  D, \;\;\; x \rightarrow \sqrt{\lambda}x \\
t &\rightarrow \mu t, \;\;\; \beta \rightarrow \mu \beta, \;\;\; D \rightarrow \mu^{-1} D, \;\;\; x \rightarrow x
\end{align}\end{subequations}
Any analytic continuation of $\tau_{1,2,3,4}$ to real time must respect these rescaling symmetries.  This forces each of the two sums above to be a function of $t/\beta$ and $x^2/D_{\mathrm{c,h}}t$ alone.   Performing the resulting sum over $n$ and integrating over $p$ (which can be subtle \cite{gu}) leads to the functional form 
\begin{equation}
\frac{1}{N^2}\sum_{I,J=1}^N \langle \Phi_I^\dagger(x,t)[\Phi_I(x,t),\Phi^\dagger_J(0,0)]\Phi_J(0,0)\rangle_\beta \sim \frac{1}{N}\left[ \mathcal{A}_{\mathrm{c}}\left(\frac{t}{\beta}, \frac{|x|}{\sqrt{D_{\mathrm{c}}t}} \right) + \mathcal{A}_{\mathrm{h}}\left(\frac{t}{\beta}, \frac{|x|}{\sqrt{D_{\mathrm{h}}t}}\right) + a_1\mathrm{e}^{2\mpi T(t-|x|/v_{\mathrm{B}})}\right]   \label{eq:sykmain}
\end{equation}   
The reason that we do not keep track of the constant prefactors is that we are only interested in the functional form of the three terms of (\ref{eq:sykmain}) to exponential accuracy.    From the above formula, we observe that the hydrodynamic regime of this model occurs once $t\gtrsim \beta$ -- hence, the time scale in our diffusion bound must be $\tau \sim \beta$.

The sum of the 8 four point functions in (\ref{eq:vBineq}) must be a sum of diffusion kernels at late times, because the long time dynamics of the charged SYK chain is a pair of two diffusion modes.    Using that \cite{gu2} \begin{equation}
\frac{1}{N}G^{\mathrm{R}}_{\rho\rho}(x,t) = \frac{1}{N^2}\sum_{I,J=1}^N \langle \Phi_I^\dagger(x,t)[\Phi_I(x,t),\Phi^\dagger_J(0,0)]\Phi_J(0,0)\rangle_\beta + \text{3 similar terms} \approx \mathcal{P}_{\mathrm{c}}(x,t) \mathrm{e}^{-x^2/4D_{\mathrm{c}}t} 
\end{equation}
with $\mathcal{P}_{\mathrm{c}}$ an algebraic function when $|x|\gg 1$ and $t\gg \beta$, we obtain from the triangle inequality that at least one of the commutators in (\ref{eq:vBineq}) must obey 
\begin{equation}
\left|\frac{1}{N^2}\sum_{I,J=1}^N \langle \Phi_I^\dagger(x,t)[\Phi_I(x,t),\Phi^\dagger_J(0,0)]\Phi_J(0,0)\rangle_\beta\right| \gtrsim \frac{\mathcal{P}_{\mathrm{c}}(x,t) \mathrm{e}^{-x^2/4D_{\mathrm{c}}t} }{4}.
\end{equation}
Hence, when $t\gg \beta$, $\mathcal{A}_{\mathrm{c}} \gtrsim \exp[-x^2/4D_{\mathrm{c}}t]$, up to constants of proportionality outside the exponential.    

%Analyzing the right hand side of (\ref{eq:sykmain}) at $x=a v_{\mathrm{B}}t$, we find that $\mathcal{A}_{\mathrm{c}} \sim \exp[-a^2v_{\mathrm{B}}^2t/D_{\mathrm{c}}]$ while the chaotic term is exponentially suppressed if $a>1$.    Since this formula is only accurate for $t \gtrsim \beta$, using the relation between $D_{\mathrm{h}}$ and $v_{\mathrm{B}}$ we conclude that if $D_{\mathrm{h}} \ge D_{\mathrm{c}}$,  $\mathcal{A}_{\mathrm{c}}$ is small compared to the chaotic term, and the butterfly lightcone dominates the commutator growth in (\ref{eq:sykmain}) at late times.
%However, if $D_{\mathrm{c}}\gg D_{\mathrm{h}}$, then the contribution to (\ref{eq:sykmain}) arising from charge diffusion can be non-negligible outside the butterfly lightcone.   Indeed, the diffusive charge term will only be inside the butterfly lightcone for times $t\gtrsim \tau_* = D_{\mathrm{c}}/v_{\mathrm{B}}^2$.   As discussed  in the main text, in this limit the chaotic light cone where OTOCs are $\mathrm{O}(1/N)$ may significantly precede the butterfly lightcone where  exponentially fast scrambling begins.

If $D_{\mathrm{c}} \lesssim D_{\mathrm{h}}$,  then both $\mathcal{A}_{\mathrm{c}}$ and $\mathcal{A}_{\mathrm{h}}$ are exponentially suppressed outside of the butterfly light cone.  Indeed 
 \begin{equation}
\mathcal{A}_{\mathrm{c}}(v_{\mathrm{B}}t,t) \sim \exp\left[-\frac{v_{\mathrm{B}}^2 t}{D_{\mathrm{c}}}\right] \sim \exp\left[-\frac{t}{\tilde\gamma\beta}\right],
\end{equation}
with a similar formula without the factor of $\tilde\gamma$ holding for $\mathcal{A}_{\mathrm{h}}$.   Now so long as $\tilde\gamma \lesssim 1$, then once $t\gtrsim \beta$ we see that these ``diffusive" contributions to OTOCs are exponentially suppressed at the butterfly light cone.   It is as if all degrees of freedom begin to scramble before the onset of hydrodynamics.

However, if $D_{\mathrm{c}}\gg D_{\mathrm{h}}$ ($\tilde\gamma \gg 1$) then there is a period of time for $\beta \lesssim t \lesssim \tilde\gamma\beta$ where charge diffusion precedes the butterfly light cone.  In this case, the light cone velocity, which sets the region of spacetime where the OTOC is at least $1/N$, will be time-dependent: \begin{equation}
v_{\mathrm{LC}}(t) = \max\left(\sqrt{\frac{D_{\mathrm{c}}}{t}},v_{\mathrm{B}}\right).
\end{equation}
In this case, we see that the charge diffusion ``light cone" can precede the butterfly light cone at short time scales.  We would find that $D_{\mathrm{c}}$ obeys the bound (\ref{eq:either}) but not (\ref{eq:mainD}), as $\tau\sim\beta$.   This effect is also somewhat trivial in this solvable model -- the \emph{only} contributions to the OTOC are either ``hydrodynamic" or ``chaotic".

Interestingly, from the expression (\ref{eq:tildegamma}) for $\tilde\gamma$ at large $q$, it appears that $D_{\mathrm{c}} \lesssim D_{\mathrm{h}}$, and in this case the butterfly light cone never lags (far) behind diffusion.    It would be interesting to numerically confirm whether or not there are points in parameter space where this model remains maximally chaotic, and so (\ref{eq:sykmain}) holds, and where we also find $D_{\mathrm{c}} \gg D_{\mathrm{h}}$.

\section{The Butterfly Effect in Holographic Models}\label{app:holovB}

Let us revisit the holographic derivation of $v_{\mathrm{B}}$ \cite{localized}.  The set-up that they consider is slightly different from (\ref{eq:vBreal}), but we expect the physics is the same.  Let us first begin by re-thinking the OTOCs of the main text in terms of correlation functions in the thermofield double state, which is a ``purification" of the thermal density matrix in a doubled Hilbert space \cite{israel1976thermo}:  \begin{equation}
|\mathrm{TFD}_\beta\rangle \equiv \frac{1}{Z(\beta)} \sum_E \mathrm{e}^{-\beta E/2}|E_{\mathrm{L}}\rangle|E_{\mathrm{R}}\rangle \langle E_{\mathrm{L}}|\langle E_{\mathrm{R}}|.
\end{equation}    
Consider a TFD perturbed by a Hermitian operator $V_{\mathrm{L}}(x,t)$, so that its quantum state is \begin{equation}
|\Psi\rangle = V_{\mathrm{L}}(x,t)|\mathrm{TFD}_\beta\rangle.
\end{equation}
Now we measure the two point function of a Hermitian operator $W$: one measured in the L theory, and the other in the R theory.   Operators in the left and right theories are defined as $W_{\mathrm{L}} = W^{\mathsf{T}} \otimes 1$ and $W_{\mathrm{R}} = 1\otimes W$.   Using the definitions above, we find 
 \begin{equation}
 \langle \Psi| W_{\mathrm{L}} (0,0)W_{\mathrm{R}}(0,0)|\Psi\rangle = \left\langle V(x,t) W(0,0) V(x,t) W\left(0,\frac{\mathrm{i}\beta}{2}\right)\right\rangle_\beta,  \label{eq:Psicor}
 \end{equation}
 with the correlation function on the right hand side now taken in the original Hilbert space.   Up to the fact that one operator is shifted halfway around the imaginary time circle, we find the OTOC half of (\ref{eq:vBreal}).   For future convenience, we also note that by considering \begin{equation}
 \left\langle \Psi \left| W_{\mathrm{R}}\left(0,-\frac{\mathrm{i}\beta}{2}\right)W_{\mathrm{R}}\left(0,0\right)\right|\Psi\right\rangle = \left\langle  V(x,t)V(x,t) W(0,0)W\left(0,\frac{\mathrm{i}\beta}{2}\right)   \right\rangle_\beta 
 \end{equation}
 we can compute a similar time-ordered correlation function.  Note also that time runs ``backwards" in theory L.
 
 The holographic dual of the TFD state is the maximally extended (two-sided) black hole at temperature $1/\beta$ \cite{maldacenaTFD}.  To compute commutators analogous to (\ref{eq:vBreal}), we should then holographically compute \begin{equation}
 \left\langle \Psi \left| W_{\mathrm{L}}(0)W_{\mathrm{R}}(0) - W_{\mathrm{R}}\left(-\frac{\mathrm{i}\beta}{2}\right) W_{\mathrm{R}}(0)\right|\Psi\right\rangle =  \tilde{h}(x,t).  \label{eq:tildeh}
 \end{equation}
 In order to compute this correlation function, we need to (\emph{i}) understand the geometry dual to $|\Psi\rangle$, and (\emph{ii}) solve bulk equations of motion for the correlation function.   We will address the former point in the next two paragraphs;  the latter point can be simplified by noting that for an operator of dimension $\Delta \gg 1$: \begin{equation}
 \langle W(x)W(y)\rangle \sim  \mathrm{e}^{-\Delta L(x,y)}  \label{eq:DeltaL}
 \end{equation}
 where $L(x,y)$ is the length of the minimal geodesic connecting the boundary points $x$ and $y$ \cite{lucasrmp}.  If the operators are located in the L/R side of the TFD Hilbert space, then the boundary points are located on the L/R boundaries of the two-sided black hole.    Note that as in (\ref{eq:vBreal}), we expect $\tilde{h}$ to have a prefactor of $1/N$.     An immediate implication of this is that,  if $|\Psi\rangle$ was in fact simply $|\mathrm{TFD}\rangle$, that the two two-point functions in (\ref{eq:tildeh}) must be equivalent when $t=0$ and $|x|$ is sufficiently large.   This equivalence can be understood by noting that the two-sided black hole geometry is a particular analytic continuation of the Euclidean black hole geometry to real time (see e.g. \cite{hartman2013time}).
 
 The geometry dual to $|\Psi\rangle$ has been argued \cite{bhbutterfly, localized} to correspond to the geometry that results from throwing a small amount of localized dust into the left side of the black hole at time $t$ in the past.    
 If $t$ is very far in the past, then this excitation will fall very close to the horizon and approximately travel along a null geodesic.  Furthermore, the stress tensor associated to this infalling matter becomes exponentially blueshifted.  As a consequence, in Kruskal coordinates (we follow the conventions of \cite{julia}), one finds that \cite{localized, blakeB1, lrbutterfly, curvedshock} \begin{equation}
\mathrm{d}s^2 = 2A\mathrm{d}U\mathrm{d}V + B\mathrm{d}\mathbf{x}^2 -2A  h(x)\mdelta(U) \mathrm{d}U^2 ,
\end{equation}   
where $U$ and $V$ are conventional null Kruskal coordinates and $\mathbf{x}$ are the boundary spatial coordinates, and  \begin{equation}
h(x) \sim G_{\mathrm{N}}  \exp\left[\frac{t}{\tau_{\mathrm{L}}} - \mu |x| \right].  \label{eq:hx}
\end{equation}
$G_{\mathrm{N}}$ is the bulk gravitational constant and is analogous to $1/N$ in the main text \cite{lucasrmp}.  The formula for $\mu$ is related to the near-horizon geometry and is unimportant for our purposes.   \cite{bhbutterfly} has computed the effect of (\ref{eq:hx}) on the geodesic length, and one finds that $\tilde{h}(x,t)\sim h(x,t)$.  Intuitively, the geodesics will be quite similar, except that the LR geodesic must also traverse the shock wave.   From (\ref{eq:tildeh}) and (\ref{eq:DeltaL}), we conclude that such a small additive change to the length of the geodesic will be proportional to $h$.   While there is a small factor of $G_{\mathrm{N}} \sim 1/N$ in front of $h$, \cite{bhbutterfly} has emphasized that this formula is valid until $t \sim t_{\mathrm{s}} \sim \tau_{\mathrm{L}} \log N$.  Hence, we conclude that \begin{equation}
v_{\mathrm{B}} = \frac{1}{\mu\tau_{\mathrm{L}}}  \label{eq:vBmu}
\end{equation}
is the butterfly velocity, as defined in the main text as the velocity with which the region where many degrees of freedom have scrambled expands outwards.

Implicit in the above argument is the assumption that the infalling matter does not backreact on the geometry beyond the small gravitational shock wave.   However, a small amount of infalling matter \emph{can} modify the geometry in other important ways.    As an example, suppose that the infalling matter has a small amount of energy density associated with it.   In the dual field theory, we would expect that a small injection of energy leads to hydrodynamic fluctuations at long time and length scales.  Indeed, in the bulk description, these hydrodynamic fluctuations are associated with metric fluctuations of strength $\mathrm{O}(1/N)$ relative to the background.  By causality, these metric fluctuations will only be non-vanishing in the left and future quadrants of the maximally extended black hole.  By the fluid-gravity correspondence \cite{flugrav1}, the metric perturbations dual to sound waves are regular and do not lead to singular perturbations near the horizon.    We thus conclude that while $\langle \Psi | W_{\mathrm{R}}(-\mathrm{i}\beta/2)W_{\mathrm{R}}(0)|\Psi\rangle$ is unaffected by both the sound wave and the shock wave,  $\langle \Psi|W_{\mathrm{L}}(0)W_{\mathrm{R}}(0)|\Psi\rangle$ can be affected by both the shock wave \emph{and} the sound wave.   We can further estimate that in the presence of a sound wave in the left half of the black hole:  \begin{equation}
\tilde{h}(x,t) \sim h(x,t) + \tilde{h}_{\mathrm{s}}(x,t)
\end{equation}
where $\tilde{h}_{\mathrm{s}}(x,t)$ is proportional to the change in the geodesic length that arises due to the fact that the left-half's metric is fluctuating in a sound wave.   The geodesic connecting the points $(x,t)=(0,0)$ in the L/R sides of the black hole will try to stay near $x=0$ and $t=0$ as much as possible.   When $t$ is large, the geodesic will only traverse the sound waves in the left half of the geometry when $|x|\ge v_{\mathrm{s}}t$.  Thus, we conclude that generically $\tilde{h}_{\mathrm{s}}(x,t) \sim 1/N$ is non-vanishing when $|x| \le v_{\mathrm{s}}t$.  Unlike $h(x,t)$, however, $h_{\mathrm{s}}(x,t)$ stays bounded and $\mathrm{O}(1/N)$ because the sound fluctuations in the metric are not singular. 

The fact that the commutator (\ref{eq:tildeh}) has two apparent contributions: one arising from the $\mathrm{O}(1/N)$ sound wave in the left geometry, and one from the gravitational shock wave at the horizon, appears quite analogous to the distinction between the light cone velocity (\ref{eq:vB}) and the butterfly velocity (\ref{eq:vBreal}) from the main text.  More carefully, we observe that if $v_{\mathrm{B}} > v_{\mathrm{s}}$, then from  (\ref{eq:vB}) we  have $v_{\mathrm{B}} = v_{\mathrm{LC}}$, with both given by (\ref{eq:vBmu}).  However, if $v_{\mathrm{s}}>v_{\mathrm{B}}$, we conclude from the argument of the previous paragraph that $v_{\mathrm{B}}$ is given by (\ref{eq:vBmu}) while $v_{\mathrm{LC}}=v_{\mathrm{s}}$.  Thus we find (\ref{eq:VLC}).

%For more complicated operators,  the bulk dynamics might include a small fluctuating sound wave \emph{and} the ``emission" of infalling matter within the sound lightcone as the sound wave passes by.   Infalling matter ``deposited" by such a sound wave at point $y$ would itself be exponentially blueshifted close to the horizon, but by the blueshift $\exp[ (t-|x-y|/v_{\mathrm{s}})/\tau_{\mathrm{L}}]$ (because the sound wave must propagate from $x$ to $y$ first).   This would imply 
%\begin{equation}
%h(x) \sim \int \mathrm{d}^dy \;  G_{\mathrm{N}} \exp\left[\frac{v_{\mathrm{s}}t  -|x-y|}{v_{\mathrm{s}}\tau_{\mathrm{L}}}\right] \mathrm{e}^{-\mu|y|} \sim G_{\mathrm{N}} \times \left\lbrace \begin{array}{ll}  \mathrm{e}^{(v_{\mathrm{B}}t-|x|)/v_{\mathrm{B}}\tau_{\mathrm{L}}} &\   |x| \le v_{\mathrm{s}}t \\  \mathrm{e}^{- (|x|-v_{\mathrm{B}}t)/v_{\mathrm{B}}\tau_{\mathrm{L}}}  &\ |x| > v_{\mathrm{s}}t \end{array}\right.\label{eq:Tsound}
%\end{equation}
%with $v_{\mathrm{B}} = \max(v_{\mathrm{s}},v_{\mathrm{B}})$.  Whether or not the stronger postulate (\ref{eq:Tsound}) is true, or there is only a small blip in the OTOC  at the sound lightcone, consistency with (\ref{eq:vB}) will indeed require (\ref{eq:mainv}).    (\ref{eq:Tsound}) may only be observed at either higher orders in a $1/N$ expansion (where, as we showed in the main text, it plays no role in (\ref{eq:either})),  or for more complicated operators.  

There is one subtle, but possibly important, difference between (\ref{eq:vBreal}) and (\ref{eq:Psicor}).   In the main text, we were interested in studying the OTOCs of non-singlet operators.   This is slightly subtle in holographic models, where one should only compute gauge-invariant quantities.   Nevertheless, we conjecture that  $V$ and $W$ in (\ref{eq:Psicor}) can be thought of as analogous to the $\Phi_I$ in (\ref{eq:vBineq}).

Another interesting interpretation of a velocity scale $\tilde{v}_{\mathrm{B}}$ has recently been put forth \cite{mezei, yang}.  If one inserts a bulk operator at a bulk spacetime point,  the dual operator $\mathcal{O}(x)$ in the field theory can be ``reconstructed" using only the degrees of freedom in a spatially localized part of the boundary theory \cite{czech, rosenhaus, wall}.   $\tilde{v}_{\mathrm{B}}$ appears to be the rate at which this reconstructable region expands as an operator moves towards the black hole at boundary spatial position $x=0$, at the speed of light \cite{mezei, yang}.  Intuitively, this is consistent with our understanding that $\tilde{v}_{\mathrm{B}}$ measures the speed at which the support of the dual operator $\mathcal{O}(x,t)$ grows with time $t$.  If the infalling matter does not backreact on the geometry, one finds $\tilde{v}_{\mathrm{B}} = v_{\mathrm{B}}$, with both given by (\ref{eq:vBmu}).   However, the dynamics becomes more complicated if the infalling matter excites a small sound wave, associated with metric perturbations of $\mathrm{O}(1/N)$  amplitude relative to the background.   Due to the background metric perturbations in the region $|x| \lesssim v_{\mathrm{s}}t$, we expect that any reconstruction procedure should include all bulk points perturbed by the sound wave.   Assuming that $\tilde{v}_{\mathrm{B}} = v_{\mathrm{LC}}$, we recover (\ref{eq:VLC}).

\section{Diffusion in Charge-Neutral Holographic Models}\label{app:holoC}
In this appendix, we discuss the theory of charge diffusion in holographic models, and derive (\ref{eq:mainD}) for these theories.   Recent reviews of holographic approaches to quantum matter may be found in \cite{lucasrmp, koenraadbook}, and for the remainder of this appendix we assume familiarity with the approach.   Our main results,  which do not require an understanding of these details, may be found in the main text.

\subsection{Geometry}
We will study large $N$ QFTs which admit a holographic dual in an asymptotically-AdS geometry.   This implies that the resulting QFT is a conformal field theory,  usually deformed by (relevant) operators in the IR.   For simplicity, we will assume, as in the main text, that the background theory is charge neutral.   

We suppose that the bulk action is given by \begin{equation}
S = S_0[g,\Phi,\ldots] - \int \mathrm{d}^{d+2}x \sqrt{-g} Z \frac{F^2}{4e^2}.  \label{eq:genbulk}
\end{equation}
The first term $S_0$ contains the action for gravity, along with additional matter content, such as scalar fields $\Phi$.   The second term consists of a decoupled Maxwell term, up to a function $Z$ which will generally depend on matter fields such as $\Phi$.   The precise form of this coupling will not be relevant for us, as will become clear.  The gauge field $A$ in the bulk is dual to a conserved current in the boundary theory, and we will compute the diffusion constant of this conserved current.   We assume that there exists an isotropic, spacetime translation invariant saddle point of the  action above, with \begin{equation}
\mathrm{d}s^2 = \frac{\mathrm{d}r^2}{U(r)} - U(r) \mathrm{d}t^2  + V(r) \mathrm{d}\mathbf{x}^2 \label{eq:metric}
\end{equation}
and $A=0$.    $r$ denotes the bulk radial coordinate and obeys $r_{\mathrm{h}}< r<\infty$.  At $r=\infty$, we have an asymptotically AdS region of the geometry where \begin{equation}
U(r\rightarrow \infty) = V(r\rightarrow\infty) = r^2 + \mathrm{O}(r).  \label{eq:asymptAdS}
\end{equation}
As $r\rightarrow 0$ we find a finite temperature event horizon, and \begin{equation}
U(r) = 4\mpi T r+ \mathrm{O}\left(r^2\right).  \label{eq:Uhor}
\end{equation}
In contrast, $V(r)$ will be finite and regular at $r=0$.  We have used the freedom to shift $r$ by a constant to fix the horizon location to $r=0$.

Assuming that we have chosen a reasonable matter action in (\ref{eq:genbulk}), the null energy condition will be obeyed in the bulk geometry.   This leads us to the following two inequalities:  \begin{subequations}\label{eq:NEC}\begin{align}
\partial_r^2 \sqrt{V} &\le 0, \\
\partial_r \left(V^{1+d/2} \partial_r \left(\frac{U}{V}\right)\right) &\ge 0.  \label{eq:NECb}
\end{align}
\end{subequations}
In particular, using (\ref{eq:Uhor}), the latter inequality  implies that \begin{equation}
U^{\prime\prime}(0) \ge -\frac{4\mpi T (d-2) V^\prime(0)}{2V(0)}.  \label{eq:Uppbound}
\end{equation}
Finally, we note that the butterfly velocity, as conventionally computed in holography, is \cite{blakeB1, lrbutterfly, curvedshock} \begin{equation}
v_{\mathrm{B}}^2  = \frac{4\mpi T}{dV^\prime(0)}.  \label{eq:vBhor}
\end{equation}
%As we describe in more detail in Appendix \ref{app:holovB}, this equation only holds for matter that has no overlap with the operators sourcing the background geometry.  We expect that this includes the operators which make up $\rho$, in the cases we are studying, up to the caveats discussed around (\ref{eq:rhoproblem}).

\subsection{The Diffusion Pole}\label{app:C2}
Having exhausted our knowledge of the background geometry, it is now time to study the two-point function of the density operator.  In particular, we will focus on finding the poles of $G^{\mathrm{R}}_{\rho\rho}$.  In the setup described above, this is equivalent to finding quasinormal modes (QNMs) of the Maxwell action.   If we find a QNM when \begin{equation}
\omega = -\mathrm{i}Dk^2 - \mathrm{i}r_4k^4 - \mathrm{i}r_6k^6 - \mathrm{i}r_8k^8 - \cdots, 
\end{equation}
this tells  us that the higher  derivative corrections to (\ref{eq:fick}) are \begin{equation}
\partial_t \rho - D\nabla^2 \rho = r_4 \nabla^4 \rho - r_6 \nabla^6 \rho + \cdots.  \label{eq:omegataylor}
\end{equation}

We now describe how to find these QNMs.   They are solutions to the bulk Maxwell equations  \begin{equation}
\partial_a \left(Z\sqrt{-g}g^{ab}g^{cd}F_{bd}\right) = 0
\end{equation}
which are both regular at $r=r_{\mathrm{h}}$, and obey $A_a(r=\infty) =0$.   We work in the gauge $A_r=0$, and look for solutions with spacetime dependence $\sim \mathrm{e}^{\mathrm{i}(kx-\omega t)}$ and $A_x$ and $A_t$ non-vanishing.   This leads to the equations \begin{subequations}\begin{align}
\left(Z\sqrt{-g}g^{rr}g^{xx} A_x^\prime\right)^\prime &= \omega Z\sqrt{-g} g^{tt}g^{xx} (\omega A_x + kA_t), \\
\left(Z\sqrt{-g}g^{rr}g^{tt} A_t^\prime\right)^\prime &= -k Z\sqrt{-g} g^{tt}g^{xx} (\omega A_x + kA_t), \\
Z\sqrt{-g}g^{rr}\left(g^{xx}k A_x^\prime - g^{tt}\omega A_t^\prime\right) &= 0.
\end{align}\end{subequations} 
In terms of the ``gauge invariant" variable \cite{lucasrmp} \begin{equation}
a = \omega A_x + kA_t,
\end{equation} 
we find the single second order differential equation \begin{equation}
\left(Z\sqrt{-g} g^{rr} g^{tt} g^{xx} \frac{a^\prime}{k^2 g^{xx} + \omega^2 g^{tt}}\right)^\prime = Z\sqrt{-g} g^{tt}g^{xx}a.
\end{equation}
In the coordinates (\ref{eq:metric}) we find \begin{equation}
\left(ZV^{d/2} \frac{a^\prime}{k^2 -\omega^2 V/U}\right)^\prime = \frac{ZV^{d/2-1}}{U}a.   \label{eq:genaeq}
\end{equation}

Unfortunately, solving (\ref{eq:genaeq})  is generally impossible, and analytic results are often available in only the simplest of theories.   Luckily, we know from (\ref{eq:omegataylor}) that there will be QNMs parametrically close to the real axis as $k\rightarrow 0$:  these correspond to the diffusion pole.   So we will solve (\ref{eq:genaeq}) perturbatively in $k$ and $\omega \sim k^2$.  It is important, however, to treat the singular near-horizon limit of (\ref{eq:genaeq}) carefully.   We write
\begin{equation}
a(r) = \left(\frac{U}{V}\right)^{-\mathrm{i}\omega/4\mpi T} s(r)
\end{equation}
with $s(\infty)=0$ and $s$ regular near $r=0$.

At leading order, (\ref{eq:genaeq}) simplifies to \begin{equation}
\left(ZV^{d/2} \frac{s^\prime}{k^2}\right)^\prime = 0,   \label{eq:km21}
\end{equation}
which has regular solutions for any $k$.  We normalize them to be
\begin{equation}
s_0 = -\int\limits_r^\infty \frac{\mathrm{d}r}{ZV^{d/2}}. \label{eq:km22}
\end{equation}
Of course, not every solution $s(r)$ of the above equations is a quasinormal mode.   We must first find a solution where $s(\infty)=0$, and where $s(r)$ is regular at the horizon $r=0$.  Plugging in the ansatz $s=s^{(0)}+s^{(1)}r + s^{(2)}r^2 + \cdots$ into (\ref{eq:genaeq}), we find \begin{align}
\omega \left(1-\frac{\mathrm{i}\omega}{2\mpi T}\right) s^\prime(0) &=\left[ \frac{\mathrm{i}k^2}{V(0)}- \frac{\mathrm{i}\omega^2 V^\prime(0)}{2\mpi TV(0)} \right] \left(1-\frac{\mathrm{i}\omega}{4\mpi T}\right) s(0)+ \frac{\mathrm{i}\omega^2 U^{\prime\prime}(0)}{(4\mpi T)^2}\left(1-\frac{\mathrm{i}\omega}{2\mpi T}\right)s(0)   \notag \\
&\;\;\;\;\;\;\;+\frac{\mathrm{i}\omega^2}{4\mpi T} \frac{(ZV^{d/2})^\prime(0)}{(ZV^{d/2})(0)} s(0)  \label{eq:regularity}
\end{align} 
Note that (\ref{eq:regularity}) is not perturbative in $\omega$ or $k$.   We simply plug in the ansatz (\ref{eq:km22}) into (\ref{eq:regularity}) and we obtain, to leading order in $k$, \begin{equation}
\omega = -\mathrm{i}Dk^2, 
\end{equation}
where \begin{equation}
D = Z(0)V(0)^{d/2-1} \int\limits_0^\infty \frac{\mathrm{d}r}{ZV^{d/2}}.
\end{equation}
This result agrees with \cite{blakeB1}, which obtained it in a different way using the Einstein relations.

We now write \begin{equation}
s(r) = s_0(r) + k^2 s_2(r) + k^4 s_4(r) + \cdots
\end{equation}
with $s_{2n}(r) \sim k^{2n}$ perturbative corrections to $s_0(r)$, as given in (\ref{eq:km22}).   When doing this perturbative expansion, we have assumed that $\omega=\omega(k)$ has been fixed by (\ref{eq:omegataylor}).   To compute $s_{2n}$, we use (\ref{eq:genaeq}) to find the following exact relation for $s$: \begin{align}
s^\prime &= \frac{\mathrm{i}\omega}{4\mpi T} \frac{V}{U}\left(\frac{U}{V}\right)^\prime s + \frac{1}{ZV^{d/2}} \left(1 - \frac{\omega^2}{k^2}\frac{V}{U}\right) \left(\frac{U}{V}\right)^{\mathrm{i}\omega/4\mpi T} \notag \\
&\;\;\;- \frac{1}{ZV^{d/2}}\left(k^2 - \omega^2 \frac{V}{U}\right) \left(\frac{U}{V}\right)^{\mathrm{i}\omega/4\mpi T} \int\limits_r^\infty \mathrm{d}r \left(\frac{U}{V}\right)^{-\mathrm{i}\omega/4\mpi T}  \frac{ZV^{d/2-1}}{U}s \label{eq:recursive}
\end{align}  
This complicated formula is useful because it is now straightforward to compute perturbative corrections to $s(0)$ and $s^\prime(0)$.   Note that at leading order, this equation reduces to (\ref{eq:km22}); alternatively, (\ref{eq:km22}) has been used to fix a constant of integration.    

Because (\ref{eq:regularity}) is exact, we need to only compute corrections to $s^\prime(0)$ and $s(0)$, order by order in $k$.   We now observe from the form of (\ref{eq:recursive}) that $s_2(0)$, $s_4(0)$, etc. will be very non-generic.  They depend on integrating a complicated function over the entire bulk radial direction.   However, some terms in the expression for $s^\prime(0)$ are ``universal" and depend only on the near horizon geometry, while others are ``non-universal" and depend on the entire geometry, far from the horizon.   For example, we may write \begin{equation}
s_2^\prime(0) =  -\frac{s_0(0) }{4\mpi T V(0)} + \frac{ s_0(0)^2 s_0^{\prime\prime}(0)}{4\mpi Ts_0^\prime(0)^2 V(0)} - \frac{ U^{\prime\prime}(0)  s_0(0)^2}{(4\mpi T)^2 s_0^\prime(0)V(0)} + \frac{ s_0(0)^2 V^\prime(0)}{2\mpi T s_0^\prime(0)V(0)^2} +  \frac{ k^2 c_2 }{Z(0)V(0)^{d/2}}  \label{eq:s2p0}
\end{equation}
where \begin{equation}
c_2 = \lim_{r\rightarrow 0} \left[ \frac{D}{4\mpi T}\log \frac{U}{V} -  \int\limits_r^\infty \mathrm{d}r \frac{ZV^{d/2-1}}{U} s_0\right]
\end{equation}
 is a constant that depends on the entire geometry.  The remaining terms in (\ref{eq:s2p0}) depend only on the fields near the horizon.   Thus, we now make a postulate that \emph{order-by-order in the $k$-expansion, the far-from-horizon terms that depend on the entire geometry will not exactly terms that were computed at subleading orders in the $k$-expansion.}   We will give a very explicit example of what we mean by this in the next paragraph.   We will see that the near-horizon regime ends up controlling the growth of $|r_4|$, $|r_6|$ etc. as defined in (\ref{eq:omegataylor}).   
 %For example, we may write that $r_6 = r_6^{\mathrm{near}} + r_6^{\mathrm{far}}$ where $r_6^{\mathrm{far}}$ is the complicated integration constant and    Unless $r_6^{\mathrm{near}} \approx -r_6^{\mathrm{far}}$,  then $|r_6| \gtrsim |r_6^{\mathrm{near}}|$ from the triangle inequality.   This approximate inequality is our postulate.   We now proceed to compute $r_n^{\mathrm{near}}$.  

Combining (\ref{eq:s2p0}) with (\ref{eq:regularity}), we obtain the entire coefficient of the quartic term in (\ref{eq:omegataylor}): \begin{equation}
 r_4 = \frac{D}{s_0^\prime(0)} c_2  - \frac{s_2(0)}{s_0^\prime(0) V(0)}.  \label{eq:r4exact}
 \end{equation}
 Rather disappointingly, there are no terms in $r_4$ that depend only on the near-horizon physics.   From the form of (\ref{eq:s2p0}), we conclude that the reason  is that the coefficients of (\ref{eq:s2p0}) have exactly cancelled with subleading corrections to the dispersion relation induced by (\ref{eq:regularity}).    The postulate of the previous paragraph is that \begin{equation}
 |r_4| \gtrsim \max\left(\frac{Dc_2}{s_0^\prime(0)} , \frac{s_2(0)}{s_0^\prime(0) V(0)}\right);  \label{eq:r4max}
 \end{equation}
 namely, $s_2(0)$ will not cancel off most of the remaining term.
 
 The cancellations between most of the near horizon terms in (\ref{eq:s2p0}) and the corrections to the regularity constraint (\ref{eq:regularity}) do not persist to higher orders. 
%  In particular, noting that\footnote{Any terms with a logarithm must not be included in the near horizon terms -- these logarithms contribute additive constants that are always sensitive to UV.} \begin{equation}
%\left. s^\prime(0)\right|_{\text{near}} = r^0\text{ term in the Taylor expansion of } \left[\frac{\mathrm{i}\omega}{4\mpi T} \frac{V}{U} \left(\frac{U}{V}\right)s + \frac{1}{ZV^{d/2}} \left(1-\frac{\omega^2}{k^2}\frac{V}{U}\right)\right],
% \end{equation}
% we can straightforwardly carry out the computation of the near horizon contributions to $r_6$, as well as higher orders.  
 At next subleading order we find \begin{align}
 r_6 &= D^3 \left(\frac{V^\prime(0)}{4\mpi T} - \frac{U^{\prime\prime}(0)V(0)}{2(4\mpi T)^2}\right)\frac{c_2}{s_0^\prime(0)} - \frac{s_4(0)}{s_0^\prime(0)V(0)} + \frac{c_2s_2(0)}{s_0^\prime(0)^2 V(0)} + \frac{Dc_4}{s_0^\prime(0)} - \frac{D^3}{(4\mpi T)^2} + \frac{D^4 s_0^{\prime\prime}(0)V(0)}{(4\mpi T)^2 s_0^\prime(0)}  \notag \\
 &+\frac{D^4 U^{\prime\prime}(0) V(0)}{(4\mpi T)^3} - \frac{D^3 c_2 V(0) s_0^{\prime\prime}(0)}{4\mpi Ts_0^\prime(0)^2} - \frac{c_2^2 D}{s_0^\prime(0)^2} \label{eq:r6exact}
 \end{align}
 where $c_4$ is a far-from-horizon constant, defined analogous to $c_2$.   We find a very long expression for $r_8$ (even the near-horizon terms only!), the $\mathrm{O}(k^8)$ term in (\ref{eq:omegataylor}).  A few of the terms in this expression are \begin{equation}
r_8 = 2D^5 \left(\frac{V^\prime(0)}{4\mpi T} - \frac{U^{\prime\prime}(0)V(0)}{2(4\mpi T)^2}\right)^2\frac{c_2}{s_0^\prime(0)} + \frac{3c_2 s_2(0) D^2}{s_0^\prime(0)^2 V(0)} \left( \frac{V^\prime(0)}{4\mpi T} - \frac{U^{\prime\prime}(0)V(0)}{2(4\mpi T)^2} \right)  + \cdots ,
\end{equation}
Now, from (\ref{eq:Uppbound}) and (\ref{eq:vBhor}) we observe that \begin{equation}
 \frac{V^\prime(0)}{4\mpi T} - \frac{U^{\prime\prime}(0)V(0)}{2(4\mpi T)^2} \le \left(1+\frac{2}{d}\right)\frac{1}{2v_{\mathrm{B}}^2}. \label{eq:r6}
 \end{equation}
 In many cases, such as the scaling theories studied in \cite{blakeB1}, this inequality is saturated.   Now, we observe the emergence of a very simple pattern arising between $r_4$, $r_6$ and $r_8$: \begin{equation}
\frac{r_6}{r_4} \gtrsim \frac{D^2}{v_{\mathrm{B}}^2}, \;\;\;\text{and/or}\;\;\; \frac{r_8}{r_6} \gtrsim \frac{D^2}{v_{\mathrm{B}}^2}, \ldots.   \label{eq:r6r4}
\end{equation}
 Indeed, at higher orders in the gradient expansion, terms in $r_{n+2}$ are proportional to terms in $r_n$, multiplied by $D^2/v_{\mathrm{B}}^2$.  For example, if the first term of (\ref{eq:r4max}) is the largest,  then using the postulate above we obtain (\ref{eq:r6r4}).   If the second term of (\ref{eq:r4max}) is the largest, then we may only conclude that $|r_8|/|r_6| \gtrsim D^2/v_{\mathrm{B}}^2$.  Nevertheless, the emergent pattern remains clear.
 
Given the UV sensitivity of almost every term in (\ref{eq:r6}), it seems unlikely to find parametric cancellations that lead to violations of (\ref{eq:r6r4}) in generic holographic models -- though we do not present an explicit proof of this statement.   We emphasize that our model has made \emph{no assumptions} about the nature of the charge neutral matter that supports the metric (\ref{eq:metric}).   Merely by demanding consistency with (\ref{eq:NEC}) and the near and far from horizon asymptotics, we can find a ``bottom up" holographic model with any $Z(r)$, $U(r)$ and $V(r)$ we want, and generically we will not find exact cancellations between the near/far from horizon terms.

%Hence we assert, as in the main text, that the breakdown of Fick's law will arise when $k^2 \gtrsim v_{\mathrm{B}}^2/D^2$.  This leads to $\tau\sim D/v_{\mathrm{B}}^2$, saturating (\ref{eq:mainD}) up to an O(1) constant.

\subsection{Quasinormal Modes at $k=0$}
In this subsection, we show that the lifetime of quasinormal modes does \emph{not} always set the time scale in (\ref{eq:mainD}), assuming that $v=v_{\mathrm{B}}$ in (\ref{eq:mainD}), and that the butterfly velocity $v_{\mathrm{B}}$ is given by (\ref{eq:vBhor}). 
Our goal is to look for quasinormal mode solutions of (\ref{eq:genaeq}), with $a$ obeying infalling boundary conditions at the horizon and $a(\infty)=0$, and $k=0$.  Although this task is not possible to do analytically in general, we will show that the quasinormal modes can be accurately computed if they are close to the real axis.  In particular, if $\omega = -\mathrm{i}/\tau_*$ with $\tau_* T \gg 1$, then
\begin{equation}
\tau_* \approx \lim_{r\rightarrow 0}\left[Z(0)V(0)^{d/2-1} \int\limits_r^\infty \frac{\mathrm{d}r}{ZV^{d/2-1}U} - \frac{1}{4\mpi T} \log \frac{T}{r}\right]  \label{eq:QNM}
\end{equation}
The factor of $T$ inside of the logarithm is not particularly important, as this formula is only accurate to leading order in $\tau_*T \gg 1$.

To derive (\ref{eq:QNM}), consider (\ref{eq:genaeq}) at very low frequencies $\omega \ll T$, and $k=0$: \begin{equation}
\left(ZUV^{d/2-1} a^\prime\right)^\prime = -\omega^2 \frac{ZV^{d/2-1}}{U}a.  \label{eq:genaeq2}
\end{equation}
Away from the horizon, as $\omega \rightarrow 0$, the approximate solution to this equation is \begin{equation}
a = \int\limits_r^\infty \frac{\mathrm{d}r}{ZUV^{d/2-1}}.  \label{eq:aC3}
\end{equation}
Suppose that there was a quasinormal mode with $\omega$ parametrically small.   Then we expect that very close to (but not exactly at) the horizon, we could approximate $a(r)$ with (\ref{eq:aC3}).  As we saw in the previous subsection, $a(r)$ must obey infalling boundary conditions at the horizon.   To linear order in $\omega$, this is equivalent to \cite{lucas1501, lucasrmp}: \begin{equation}
a(r\rightarrow 0) \approx s_0 \left[1 + \frac{\mathrm{i}\omega}{4\mpi T} \log \frac{1}{r}\right]. \label{eq:ato0}
\end{equation}
Indeed, (\ref{eq:aC3}) has a logarithmic divergence:  \begin{equation}
a(r\rightarrow 0) \approx s_0 + \frac{1}{Z(0)V(0)^{d/2-1}} \frac{1}{4\mpi T} \log \frac{T}{r}
\end{equation}
with 
\begin{equation}
s_0 = \lim_{r\rightarrow 0} \left[\int\limits_r^\infty \frac{\mathrm{d}r}{ZUV^{d/2-1}} - \frac{1}{Z(0)V(0)^{d/2-1}} \frac{1}{4\mpi T}\log\frac{T}{r}\right]  \label{eq:s0def}
\end{equation}
is a finite non-zero constant.   Combining (\ref{eq:ato0}) then implies (\ref{eq:QNM}).   Note that $s_0$ is formally sensitive to the factor of $T$ in the logarithm.   So, as we cautioned above, we should only take (\ref{eq:QNM}) seriously to leading order in $\tau_*T \gg 1$.   If we predict $\tau_* \sim 1/T$,  then the numerical prefactor cannot be trusted, but we will have confirmed the absence of a quasinormal mode with $\tau_* \ll 1/T$, which is sufficient for our purposes.

We can estimate the breakdown of (\ref{eq:ato0}) by asking when the imaginary logarithmic term is comparable in magnitude to the real term, and we observe that this occurs only for \begin{equation}
r \lesssim T \mathrm{e}^{-T\tau_*}.
\end{equation}
Thus, when $T\tau_* \gg 1$, our approximation of $a(r)$ is valid for almost the entire geometry.   We can now use a standard holographic matching argument \cite{lucasrmp} to show that (\ref{eq:ato0}) straightforwardly transitions to a regular solution $a(r)$ obeying infalling boundary conditions at the horizon.   When \begin{equation}
r \ll \frac{4\mpi T}{|U^{\prime\prime}(0)|}
\end{equation}
 we may approximate $a(r)$ by the solution to the differential equation \begin{equation}
0 \approx  r^2 a^{\prime\prime} + r\left(1+br\right)  a^\prime + \frac{\omega^2}{(4\mpi T)^2} a,  \label{eq:anearhor}
\end{equation}
with \begin{equation}
b = \frac{(ZV^{d/2-1})^\prime(0)}{(ZV^{d/2-1})(0)},
\end{equation}
The solution to (\ref{eq:anearhor}) is \begin{equation}
a(r) = C x^{\mathrm{i}\omega/4\mpi T} \mathrm{e}^{-bx}\left[\mathrm{U}\left(-1-\frac{\mathrm{i}\omega}{4\mpi T}, 1+\frac{\mathrm{i}\omega}{2\mpi T}, bx\right) - \frac{\mathrm{\Gamma}(-\frac{\mathrm{i}\omega}{2\mpi T})}{\mathrm{\Gamma}(1-\frac{\mathrm{i}\omega}{4\mpi T})} \frac{\mathrm{L}_{-1-\mathrm{i}\omega/4\mpi T}^{\mathrm{i}\omega/2\mpi T}(bx)}{\mathrm{L}_{-1-\mathrm{i}\omega/4\mpi T}^{\mathrm{i}\omega/2\mpi T}(0)}\right],  \label{eq:Dmatch}
\end{equation}
where $C$ is an undetermined constant, $\mathrm{L}^a_b(x)$ is the generalized Laguerre polynomial, and $\mathrm{U}(a,b,x)$ is the confluent hypergeometric function of the second kind.  The linear coefficients in front of these two terms have been fixed by imposing infalling boundary conditions (thus setting the $\mathrm{O}(x^0)$ term inside the square brackets to vanish).   Taylor expanding (\ref{eq:Dmatch}) at small $\omega$, and then small $r$, we obtain \begin{equation}
a(r) = C\left[\frac{2\mpi T}{\mathrm{i}\omega} - \frac{\log r}{2} + \mathrm{O}\left(r^0,\omega^0\right) \right],
\end{equation}
which confirms that (\ref{eq:ato0}) is the approximate solution to the equation of motion in the matching region $T\mathrm{e}^{-T\tau_*} \lesssim r \lesssim T/|U^{\prime\prime}(0)| \sim T$.   This confirms that our estimate of $\tau_*$ is accurate so long as $\tau_* T \gg 1$.

Let us now give a simple example where we can clearly estimate that the location of this quasinormal mode is at a far larger frequency scale than $D/v_{\mathrm{B}}^2$.   A simple family of models which exhibits a divergence where $D/v_{\mathrm{B}}^2 \gg 1/T$ is a charge-neutral Lifshitz theory \cite{kachru, lucasrmp} with $Z=1$ and dynamical critical exponent $z>d$.  In our coordinate system, one finds that for $r\ll \Lambda$ \cite{blakeB1} \begin{subequations}\begin{align}
U(r) &= (r+r_0)^2 \left(1-\left(\frac{r_0}{r_0+r}\right)^{1+d/z}\right)   \\
V(r) &=  \Lambda^2 \left(\frac{r+r_0}{\Lambda}\right)^{2/z}
\end{align}\end{subequations}
with \begin{equation}
r_0 \equiv \frac{4\mpi Tz}{z+d}.
\end{equation}
Here $\Lambda$ is a UV energy scale, beyond which the geometry asymptotes to the Poincar\'e patch of $\mathrm{AdS}_{d+2}$; for simplicity, we will not worry about the precise completion to AdS.   We note in passing that (\ref{eq:Uppbound}) is saturated, as we claimed it would be for such a scaling theory.   We estimate that \begin{equation}
D \sim \Lambda^{d-2} \left(\frac{T}{\Lambda}\right)^{(d-2)/z} \int\limits_{r_0}^\infty \frac{\mathrm{d}r}{r^{d/z}} \Lambda^{-d+d/z} \sim  \frac{1}{\Lambda} \left(\frac{T}{\Lambda}\right)^{(d-2)/z} \sim \frac{v_{\mathrm{B}}^2}{T} \left(\frac{\Lambda}{T}\right)^{(z-d)/z},  \label{eq:Dlifshitz}
\end{equation}
where in the last step we have used that $v_{\mathrm{B}} \sim (T/\Lambda)^{1-1/z}$, which follows from (\ref{eq:vBhor}).   However we find \begin{equation}
s_0 \sim \int\limits_T^\Lambda \frac{\mathrm{d}r}{r^{2+(d-2)/z} \Lambda^{(d-2)(1-1/z)}} \sim \frac{1}{T^{1+(d-2)/z}\Lambda^{(d-2)(1-1/z)}} 
\end{equation}
which leads to a quasinormal mode at \begin{equation}
\omega \sim -\mathrm{i}T.
\end{equation}
Hence, we find that $\tau_* \ll \tau$.   In a scalar sector of these Lifshitz theories, one can  show that (\emph{i}) QNMs have not moved close to the real axis \cite{sybesma}, and (\emph{ii}) QNMs lie on the imaginary axis for $d<z$, consistent with our ansatz $\omega= -\mathrm{i}/\tau_*$.

In some respects, the claim that there is no non-quasinormal mode heralding the breakdown of hydrodynamics  is a rather surprising observation: in the simplest holographic model (Einstein gravity) \cite{janik} has observed that a Borel-resummed hydrodynamics is able to predict the locations of quasinormal modes in holography,  in a simple model.  Our estimate for the smallest quasinormal mode is consistent with the fact that we find that ``universal" terms $\sim D^3/v_{\mathrm{B}}^2$ do not arise in $r_4$.  In particular, suppose that we applied the regularity condition (\ref{eq:regularity}) to (\ref{eq:km22}), but kept $\mathrm{O}(\omega^2)$ terms in (\ref{eq:regularity}).  Then we find (up to a non-universal contribution at $\mathrm{O}(\omega^2)$): \begin{equation}
\omega \sim -\mathrm{i}D\left(k^2 - \frac{\omega^2}{v_{\mathrm{B}}^2}\right).
\end{equation}
This equation would have a quasinormal mode at $\omega \sim - \mathrm{i}v_{\mathrm{B}}^2/D$, which we do not find above.  It would be interesting to more carefully resolve this puzzle in a particular non-trivial scaling geometry.  

A final interesting observation is that $D \lesssim c^2 \tau_*$, where $c\sim v_{\mathrm{LR}}$ is the speed of light in the UV theory:\begin{equation}
\tau_* \approx Z(0)V(0)^{d/2-1} \int\limits^\infty_T \frac{\mathrm{d}r}{ZV^{d/2}} \frac{V}{U}  \gtrsim Z(0)V(0)^{d/2-1} \int \limits_0^\infty \frac{\mathrm{d}r}{ZV^{d/2}} = D
\end{equation}
where we have exploited that $V\ge U$ by the null energy condition.  To prove this inequality, we integrate (\ref{eq:NECb}) once and obtain $(U/V)^\prime > 0$ by studying the near-horizon limit.   The asymptotically AdS boundary conditions then fix $U=V$ at $r=\infty$, so we conclude $V\ge U$.    However, the bound $D \lesssim c^2 \tau_*$ is often very weak compared to $D \lesssim v_{\mathrm{B}}^2 \tau$,  and $G^{\mathrm{R}}_{\rho\rho}$ does not always take a diffusive form for times larger than $\tau_*$.  We do not know whether $D \lesssim c^2 \tau_*$ is simply a curiosity of this holographic model, or signifies a more profound result.

\end{appendix}
 
\bibliographystyle{unsrt}
\addcontentsline{toc}{section}{References}
\bibliography{inhomosykbib}

\end{document}